\documentclass[11pt,a4paper]{article}
\usepackage{jcappub}

\usepackage[utf8]{inputenc}
\usepackage[T1]{fontenc}
\usepackage{amsmath,amssymb}
\usepackage{verbatim}

\DeclareMathOperator{\sgn}{sgn}

\title{Gauge-Invariant Perturbations in Hybrid Quantum Cosmology}
\date{\today}

\author[a]{Laura Castell\'o Gomar,}

\author[b]{Mercedes Mart\'in-Benito}

\author[a]{and Guillermo A. Mena Marug\'an}

\affiliation[a]{Instituto de Estructura de la Materia, IEM-CSIC,\\ Serrano 121, 28006 Madrid, Spain}
\affiliation[b]{Radboud University Nijmegen, Institute for Mathematics, Astrophysics and Particle Physics, \\Heyendaalseweg 135, NL-6525 AJ Nijmegen, The Netherlands}

\emailAdd{laura.castello@iem.cfmac.csic.es}
\emailAdd{m.martin@hef.ru.nl}
\emailAdd{mena@iem.cfmac.csic.es}

\abstract{We consider cosmological perturbations around homogeneous and isotropic spacetimes minimally coupled to a scalar field and present a formulation which is designed to preserve covariance. We truncate the action at quadratic perturbative order and particularize our analysis to flat compact spatial sections and a field potential given by a mass term, although the formalism can be extended to other topologies and potentials. The perturbations are described in terms of Mukhanov-Sasaki gauge invariants, linear perturbative constraints, and variables canonically conjugate to them. This set is completed into a canonical one for the entire system, including the homogeneous degrees of freedom. We find the global Hamiltonian constraint of the model, in which the contribution of the homogeneous sector is corrected with a term quadratic in the perturbations, that can be identified as the Mukhanov-Sasaki Hamiltonian in our formulation. We then adopt a hybrid approach to quantize the model, combining a quantum representation of the homogeneous sector with a more standard field quantization of the perturbations. Covariance is guaranteed in this approach inasmuch as no gauge fixing is adopted. Next, we adopt a Born-Oppenheimer ansatz for physical states and show how to obtain a Schr\"odinger-like equation for the quantum evolution of the perturbations. This evolution is governed by the Mukhanov-Sasaki Hamiltonian, with the dependence on the homogeneous geometry evaluated at quantum expectation values, and with a time parameter defined also in terms of suitable expectation values on that geometry. Finally, we derive effective equations for the dynamics of the Mukhanov-Sasaki gauge invariants, that include quantum contributions, but have the same ultraviolet limit as the classical equations. They provide the master equation to extract predictions about the power spectrum of primordial scalar perturbations.}


\begin{document}
	
\maketitle
	
\section{Introduction}

The treatment of linear perturbations has become a prolific field in modern cosmology. The advent of inflationary universe models \cite{inflationmodels} generated great interest in cosmological perturbations and gave rise to a lot of activity in this area. Beyond its early success, the study of perturbed systems in general relativity requires great care because one must deal with the gauge freedom inherent to the theory, which affects the description of the perturbations. Once this problem was realized, most efforts focused on getting a manifestly gauge-invariant formalism. 

Nowadays, cosmology is entering a golden age owing to the recent observational progress, which has opened new windows to test the predictions of theoretical models \cite{precision}. The latest observations are providing us with increasingly accurate data of cosmological phenomena and, for the first time, it seems possible for astrophysics to think about finding traces of the quantum geometric structure of the early history of the universe \cite{gravpert}. For this reason, the ultimate hope of the community of physicists working in the quantization of gravity is to develop a quantum theory capable of leading to testable predictions. In order to fully capture the quantum nature of spacetime, this theory must involve simultaneously both the geometry and the perturbations, with interplay between them. 

Over the last decades, the theory of cosmological perturbations \cite{theorypertu}, combined with the inflationary paradigm \cite{inflationmodels}, has emerged as the framework which conciliates the theoretical models of the early universe with observations, since it provides a good approximation to the anisotropies of the cosmic microwave background (CMB) and explains quite satisfactorily the formation of structures at large scales \cite{inflation}. The study of the CMB is a powerful tool for understanding the universe in its origins. It supports the approximation that the observed region is homogeneous and isotropic in a suitable average (demonstrated under certain theoretical assumptions \cite{EGS}). However, this leads to questions about how the anisotropies and cosmological structures formed and developed. 

The pioneer work in the analysis of perturbations around classical Friedmann-Robertson-Walker (FRW) cosmologies\footnote{These cosmologies are also called Friedmann-Lema\^itre-Robertson-Walker cosmologies by many authors.} is due to Lifshitz \cite{Lpert}, as a first modelization of the universe considered at a large scale. Nonetheless, it was relatively soon noticed that this analysis had been carried out with a specific gauge choice, and hence it did not address the gauge freedom satisfactorily. This gauge dependence in the description of the perturbations has caused many controversies because keeping track of the gauge modes can get cumbersome, something which makes difficult the extraction of the physically meaningful degrees of freedom. An attempt to provide a covariant treatment of the perturbations was made by Hawking \cite{Hpert}, but this work did not resolve totally the gauge ambiguities. It was completed later by Olson \cite{Olson} for the case of an isentropic perfect fluid in a spatially flat spacetime. However, it was Bardeen who first constructed a truly gauge-invariant formalism (originally for a perfect fluid), which mixes the perturbations of the matter with the perturbations of the four-dimensional metric \cite{bardeen}. This work was followed by many other contributions \cite{gaugeinvariant,sasa}. Likewise, Mukhanov, based on Sasaki's investigations \cite{sasa}, proposed some gauge-invariant field-like variables for the case of a scalar field on a spatially flat FRW background, directly related to the co-moving curvature perturbations \cite{mukhanov}. Mukhanov expanded the action for the gravitational and scalar fields up to second order in the perturbations and introduced a gauge-invariant field that completely characterized those perturbations and allowed one to rewrite their action exclusively in terms of it when the background equations were employed.  

In a standard analysis of primordial fluctuations, one studies the perturbations within the scheme of Quantum Field Theory (QFT) in a classical and fixed curved spacetime. From this viewpoint, the nature of these primordial perturbations is in fact quantum and it is rather general to assume that they started out at a very early time as fluctuations with a small amplitude and gradually grew in time as consequence of gravitational instabilities. For doing this, essentially, one represents the perturbations by quantum fields and considers that they were initially propagating on a given geometry, such as a de Sitter spacetime, which describes rather well the inflationary stage of the universe \cite{bunchdavies}. Actually, we already know from the CMB data that fluctuations with amplitudes of $10^{-5}$ are sufficient to reproduce the cosmic structures observed today. Despite how well this treatment is able to explain the present observations, the challenge for quantum cosmology is to build a formalism which includes the homogeneous background and the inhomogeneous perturbations and proves to be potentially predictive, in order to elucidate whether the relics of the quantum fluctuations of the early universe may encode information about the quantum character of the spacetime geometry itself. 

A canonical approach to cosmological perturbations in which the background and the inhomogeneities (both in the geometry and in a matter scalar field) were treated together quantum mechanically, although fixing the gauge, was developed by Halliwell and Hawking \cite{HH}, in the context of closed FRW universes. They expanded the action up to second order in the perturbations and then built the corresponding Hamiltonian. Shortly after, Shirai and Wada \cite{shiwa} reformulated this formalism, isolating the quantum dependence on gauge invariants. Actually, they introduced not only a canonical transformation for the perturbations but modified the variables of the background by terms quadratic in those perturbations. Even so, physical states still depended on perturbative variables that were not gauge invariants (although in a way that was completely determined), and the introduced canonical transformations were only defined in a semiclassical limit. Therefore, they depended on the solution to the effective equations that was considered in each case. Later, Langlois \cite{Langlois} tried to refine and clarify this procedure by using techniques inspired by Hamilton-Jacobi theory in order to obtain the gauge-invariant perturbations. Nevertheless, he did not include transformations of the background and, besides, obviated the explicit form in which the Hamiltonian constraint depends on non gauge-invariant variables, because this should be physically irrelevant. More recently, Pinto-Neto and collaborators \cite{PinhoPinto} proposed another approach by means of canonical transformations which involve the perturbations and the background and, in particular, include the Mukhanov-Sasaki (MS) variables. More specifically, they considered the case of a perfect fluid and of a scalar field, and reformulated the system so that the global Hamiltonian constraint depends only on the gauge invariants and the new background variables. In this reformulation, the equations of motion of the background were not used, and the lapse function was redefined. Nonetheless, second-class constraints appeared in this reformulation procedure which had to be eliminated by reduction and the introduction of Dirac brackets, a step that obscures the full gauge invariance of the ultimate system and the role in it of the perturbative constraints. 

In the last years, the canonical quantization of cosmological perturbations has received as well a lot of attention within the framework of Loop Quantum Cosmology (LQC). LQC \cite{lqc} addresses the quantization of cosmological systems using the ideas and techniques of Loop Quantum Gravity (LQG) \cite{lqg}, a non-perturbative and background independent program for the quantization of general relativity that provides one of the most promising approaches for a quantum theory of the gravitational interaction. Initially, LQC was applied successfully to homogeneous models in cosmology, leading to a consistent quantization of the FRW spacetime in which the Big Bang singularity is replaced with a quantum bounce, namely the Big Bounce \cite{bounce1,bounce2}. Clearly, the limitation of homogeneity is a restriction that must be surpassed when one is interested in studying more realistic scenarios. Therefore, it is natural to try and go beyond this simplification and incorporate small fluctuations in the quantum treatment of the geometry and the matter content.

The loop quantization of the FRW background has been combined with a Fock quantization of the perturbations in a so-called {\it{hybrid}} approach in which the whole system that describes the perturbed cosmologies is quantized with canonical methods \cite{hybridFRW,hybridFRWflat,hybridFRWflatMS}. In particular, in this formalism the perturbed spacetime geometry can be treated as a fully quantum entity. This hybrid approach was originally developed in the linearly polarized Gowdy models \cite{hybridgowdy}. In this case, the inhomogeneities can be dealt with exactly, rather as pertubatively. An alternative proposal, put forward by Ashtekar {\it et al.} \cite{dressed}, investigates the propagation of the perturbations in the {\it{dressed}} FRW geometry obtained by incorporating the quantum effects of LQC on the background. Consequently, in this analysis, known as the {\it{dressed metric}} formalism, one loses a full quantum description of the geometry in the perturbed system. Both, in the {\it{hybrid}} and the {\it{dressed metric}} approaches in LQC, the perturbations have been expressed in terms of gauge invariants, though just after eliminating degrees of freedom in a reduction that casts doubt on the covariance of the system. Finally, a third route to analyze the effects of LQG in cosmological perturbations consists in assuming corrections in the quantum constraints arising from the use of holonomies and the regularization of the inverse of the volume operator \cite{effective,BBC}. Demanding the closure of the modified algebra of constraints, one can deduce the form of those quantum corrections and the corresponding effective field equations for the propagation of the perturbations. This strategy is intrinsically covariant, although it still rests on a series of hypotheses about, e.g., the form of the possible loop quantum corrections, the use of local expansions for non-local quantities, or the independence of the results on the existence of superselection sectors.

In this work, we will provide a canonical formalism for perturbed flat FRW spacetimes with a matter scalar field which is specially designed to preserve covariance at the considered perturbative level. In particular, the perturbations will be described by the MS gauge invariants, by combinations of the perturbative constraints, and by variables canonically conjugate to them. For completeness and self-consistency in the presentation, we will detail the change to this description from a classical formulation similar to that of  Halliwell and Hawking \cite{HH} for the case of flat spatial topology. This change will be obtained by means of a canonical transformation which will be completed in order to include as well the background variables. In addition, we will discuss how to quantize the resulting model using generalized hybrid techniques, namely, combining any given quantization of the background with a quantum theory for the perturbations (assuming the compatibility of both quantizations in the whole system). In this framework, we will show how to extract quantum field equations for the MS variables without any gauge fixing. Furthermore, this will be done without appealing to semiclassical approximations, like in Refs. \cite{HH,shiwa}, nor a Bohm-De Broglie scheme, like in Ref. \cite{PinhoPinto}. Instead, we will employ a kind of Born-Oppenheimer ansatz and discuss the validity of its application. In this derivation, we will not make use of any background equation of motion or constraint, neither at a classical, effective, nor quantum level. Finally, we will particularize our discussion to the specific case of a hybrid quantization in loop cosmology. 

In this sense, our analysis extends other treatments like that of Ref. \cite{hybridFRWflatMS}, where the gauge freedom associated with the perturbative constraints was fixed, while we do not make any such classical gauge fixing here. For that, we will need to abelianize the algebra of constraints in our perturbed FRW system. Let us finally mention that, in the context of LQG as well, there have been some attempts to develop a manifestly gauge-invariant perturbation theory in the canonical framework by constructing approximate complete observables \cite{bianca}, adopting an approach different from ours.

The plan of the rest of the paper is as follows. Sec. \ref{sec:system} contains the notation and the description of the classical perturbed system. In Sec. \ref{sec:inv} we will introduce the set of transformations for the inhomogeneous modes that leads to gauge invariants, perturbative constraints, and their conjugate variables. This will be extended in Sec. \ref{sec:complete} to a transformation for the whole perturbed system that preserves the canonical structure. In doing this, we will have to include quadratic perturbative corrections to the background variables. In addition, in this section we will study how these changes modify the zero-mode of the Hamiltonian constraint. In Sec. \ref{sec:hybrid} we will explain how to proceed to a (generalized) hybrid quantization of the total system formed by the background and the perturbations. Adopting then a Born-Oppenheimer ansatz, we will deduce a quantum equation for the propagation of the MS variables, discussing its range of validity. Besides, we will derive an effective equation governing the evolution of the corresponding classical variables. This formalism will be applied to LQC in Sec. \ref{sec:lqc}. We will conclude and summarize in Sec. \ref{sec:conclusions}. Finally, we will include two appendices with extra details of the calculations.

\section{The  perturbed flat FRW model}
\label{sec:system}

In this section we review the classical description of the FRW model with a minimally coupled scalar field and with inhomogeneous perturbations (see e.g. Ref. \cite{HH}). We consider the case of flat spatial topology which (in order to avoid technical problems in the ultraviolet regime, as we will comment later on) we assume to be compact, namely that of a three-torus. This flat case describes the physically relevant situation in cosmology, since it is in agreement with observations, as long as the compactification scales of the three-torus are large compared to the radius of our observable universe. The matter content of the model is provided by a scalar field $\Phi$ minimally coupled to the geometry and subject to a potential term. Although the discussion can be carried out for generic potentials, we will particularize our analysis to the case of a mass contribution for simplicity, explaining just how the main formulas of our study generalize for any other potential of the scalar field. On the other hand, as it is well known, scalar, vector, and tensor perturbations decouple from each other at leading order in the perturbative treatment, and may be considered independently. Actually, for scalar fields, vector perturbations are pure gauge and therefore do not contain any physical degree of freedom. On the contrary, tensor perturbations are in fact gauge invariant, and hence their treatment is relatively simple, at least with respect to the issues of covariance of the perturbative formulation that we want to address. Besides, from an observational point of view, scalar perturbations are the most interesting ones, since they are ultimately responsible for the anisotropies measured in the temperature of the CMB. We will thus focus our discussion on scalar perturbations. This model was also considered in Refs. \cite{hybridFRWflat, hybridFRWflatMS} but, unlike in this work, the system was reduced by partially fixing the gauge freedom in those references. Here, we follow conventions and notation similar to the ones of those works, and we refer the reader to them for further details. 

Adopting a canonical 3+1 decomposition, we parameterize the four-metric in terms  of the three-metric $h_{ij}$ induced on the sections of constant time $t$, a lapse function $N$, and a shift vector $N^i$ (or co-vector $N_i$). Spatial indices $i,j$ run from 1 to 3. For the unperturbed FRW spacetime, these metric functions are completely characterized by a homogeneous lapse $N_0(t)$, and by the scale factor of the spatial metric, its square multiplying a static auxiliary three-metric $^0h_{ij}$. Instead of using the (positive) scale factor we will consider its (real) logarithm, denoted by $\alpha(t)$. On the other hand, we take $^0h_{ij}$ as the standard Euclidean metric on the three-torus $T^3$, appropriate for
the considered case of a compact flat universe. We denote the corresponding angular coordinates by $\theta_i$, such that $2\pi \theta_i/ l_0 \in S^1$, so that the period is $l_0$ in each of the orthonormal directions. 

Any function defined on the spatial sections, such as those describing the inhomogeneous perturbations, can be expanded in the basis formed by the eigenmodes of the Laplace-Beltrami operator compatible with the metric  $^0h_{ij}$. In this way, we transform the study of the spatial dependence into a spectral analysis in terms of those modes, whose dynamics decouple (in fixed backgrounds) at leading perturbative order. Then, as in Ref. \cite{hybridFRWflatMS}, we adopt a basis of real Fourier modes, formed by the sine and cosine functions
\begin{eqnarray}
\tilde Q_{\vec n,+} (\vec\theta)&=& \sqrt 2\cos\left(\frac{2\pi}{l_0}\vec n\cdot\vec\theta\right),\\  \tilde Q_{\vec n,-} (\vec\theta)&=& \sqrt 2\sin\left(\frac{2\pi}{l_0}\vec n\cdot\vec\theta\right).
\end{eqnarray}
Here, $\vec n\cdot\vec\theta=\sum_in_i\theta_i$, and $\vec n=(n_1,n_2,n_3)\in\mathbb Z^3$ is any tuple of integers such that its first non-vanishing component is strictly positive (this restriction is introduced to avoid repetition of modes). These scalar modes have a norm equal to the square root of the auxiliary volume $l_0^3$ of the three-torus, and a Laplace-Beltrami eigenvalue equal to $-\omega_n^2=-4\pi^2\vec n\cdot\vec n/l_0^{2}$. 

In the expansion of the inhomogeneities, the vanishing tuple $\vec n$ is not included. The zero-modes account for the background homogeneous geometry, parameterized by $N_0(t)$ and $\alpha(t)$, and for the homogeneous part of the matter field $\Phi$, that we denote by $\varphi(t)$. These degrees of freedom are treated exactly at the perturbative order of our truncations in the description of the system.\footnote{See e.g. Refs. \cite{HH,shiwa,PinhoPinto,kk}, and the discussion on this issue in Refs. \cite{BBC, hybridFRWflatMS}, as well as the full treatment beyond perturbation theory adopted in the inhomogeneous Gowdy cosmologies \cite{gow}, that confirms that when the inhomogeneities are regarded as perturbations, this procedure to deal with the zero-modes is correct.} Using this Fourier expansion, the spacetime metric and the scalar field can be expressed as
\begin{subequations}\label{eqs:expansions}
 \begin{eqnarray}\label{3metric}
h_{ij}(t,\vec\theta) &=& \sigma^2 e^{2\alpha(t)}\;{}^0h_{ij}(\vec\theta )\left[1+2\sum_{\vec n,\epsilon}a_{\vec n,\epsilon} (t)\tilde Q_{\vec n,\epsilon}(\vec\theta)\right] \nonumber\\
&+&6 \sigma^2 e^{2\alpha(t)}\sum_{\vec n,\epsilon}b_{\vec n,\epsilon}(t)\left[\frac1{\omega_n ^2}(\tilde Q_{\vec n,\epsilon})_{|ij}(\vec\theta)+\frac13{}^0h_{ij}(\vec\theta)\tilde Q_{\vec n,\epsilon}(\vec\theta)\right],\\
\label{lapse}
N(t,\vec\theta) &=& \sigma \left[N_0(t)+e^{3\alpha(t)} \sum_{\vec n,\epsilon}g_{\vec n,\epsilon}(t)\tilde Q_{\vec n,\epsilon}(\vec\theta)\right], \\ 
\label{shift}
N_i(t,\vec\theta) &=& \sigma^2e^{2\alpha(t)}\sum_{\vec n,\epsilon}\frac1{\omega_n^2}k_{\vec n,\epsilon}(t)(\tilde Q_{\vec n,\epsilon})_{|i}(\vec\theta),\\
\label{field}
\Phi(t,\vec\theta) &=& \frac{1}{\sigma\sqrt{l_0^{3}}}\left[\varphi(t)+\sum_{\vec n,\epsilon}f_{\vec n,\epsilon}(t)\tilde Q_{\vec n,\epsilon} (\vec\theta)\right].
\end{eqnarray}
\end{subequations}
In these equations, we have defined the constant $\sigma^2=4\pi G/(3l_0^3)$, $G$ is the Newton constant, the vertical bar stands for the covariant derivative with respect to the auxiliary metric ${}^0h_{ij}$, and $\epsilon=+,-$ for cosine and sine modes, respectively. Besides, we have scaled for convenience the shift vector and the inhomogeneous part of the lapse function by a power of the scale factor\footnote{We note that ${k}_{\vec n,\epsilon}(t)$ and ${g}_{\vec n,\epsilon}(t)$ are not exactly those of Ref. \cite{hybridFRWflatMS}, owing to the commented scaling.}  (and of the mode frequency in the case of the shift). The time-dependence of the geometry and matter scalar perturbations is parameterized by the functions
\begin{equation}\label{vari}
\{a_{\vec n,\epsilon} (t), b_{\vec n,\epsilon} (t), g_{\vec n,\epsilon} (t), k_{\vec n,\epsilon} (t),  f_{\vec n,\epsilon} (t)\}.
\end{equation} 
In what follows we will omit the time-dependence in these quantities to simplify the notation.

Following an approach parallel to that in Ref. \cite{HH}, we can now substitute these expressions in the Hamiltonian form of the gravitational action with a minimally coupled scalar field and truncate the result at quadratic order in the inhomogeneous perturbations. In this way, one obtains the total Hamiltonian $H$ of the perturbed system at this order of approximation. As expected, this Hamiltonian is given by 
a linear combination of constraints, reflecting the freedom inherited from general relativity to perform spatial diffeomorphisms and time reparameterizations. Specifically, we get \cite{hybridFRWflat,HH}
\begin{equation}\label{eq:Hamiltonian}
H ={N}_0\Big[H_{|0}+\sum_{\vec n,\epsilon} H^{\vec n,\epsilon}_{|2}\Big]+\sum_{\vec n,\epsilon} g_{\vec n,\epsilon} \tilde{H}^{\vec n,\epsilon}_{|1}+\sum_{\vec n,\epsilon}k_{\vec n,\epsilon} 
\tilde{H}^{\vec n,\epsilon}_{\_1}.
\end{equation}
Here, $H_{|0}$ denotes the scalar or Hamiltonian constraint of the unperturbed flat FRW model, that generates homogeneous time reparameterizations in that system:
\begin{equation}\label{eq:H_0}
 H_{|0} = \frac{e^{-3\alpha}}{2}\big(-\pi_\alpha^2+\pi_\varphi^2+e^{6\alpha}\tilde{m}^2\varphi^2).
\end{equation}
The constant $\tilde{m}$ is the mass $m$ of the scalar field conveniently re-expressed as $\tilde{m}= m \sigma$. Besides, we employ the notation $\pi_q$ to denote the momentum conjugate to any variable $q$. Notice that,
in the perturbed case under study, the zero-mode of the Hamiltonian constraint (or global Hamiltonian constraint) gets contributions from the inhomogeneities. At our truncation order, these contributions are quadratic in the perturbations. We have included them in the sum of terms  $ H^{\vec n,\epsilon}_{|2}$. For each mode of the perturbations, the contribution is\footnote{This formula corrects a misprint in Ref. \cite{hybridFRWflat}, that was not relevant for the results discussed in that work.} 
\begin{align}\label{eq:H2}
H^{\vec n,\epsilon}_{|2} &= \frac{e^{-3\alpha}}{2}\Big\{ -\pi_{a_{\vec n,\epsilon}}^2+\pi_{b_{\vec n,\epsilon}}^2+\pi_{f_{\vec n,\epsilon}}^2+2\pi_\alpha(a_{\vec n,\epsilon}\pi_{a_{\vec n,\epsilon}}+4b_{\vec n,\epsilon}\pi_{b_{\vec n,\epsilon}})-6\pi_\varphi a_{\vec n,\epsilon}\pi_{f_{\vec n,\epsilon}} \nonumber\\
&\phantom{=\frac12e^{-3\alpha}\Big\{} +\pi_\alpha^2\Big(\tfrac12a_{\vec n,\epsilon}^2+10b_{\vec n,\epsilon}^2\Big)+\pi_\varphi^2\Big(\tfrac{15}2a_{\vec n,\epsilon}^2+6b_{\vec n,\epsilon}^2\Big) \nonumber\\
&\phantom{=\frac12e^{-3\alpha}\Big\{} -e^{4\alpha}\Big(\tfrac13\omega_n^2a_{\vec n,\epsilon}^2+\tfrac13\omega_n^2b_{\vec n,\epsilon}^2+\tfrac23\omega_n^2a_{\vec n,\epsilon}b_{\vec n,\epsilon}-\omega_n^2f_{\vec n,\epsilon}^2\Big) \nonumber\\
&\phantom{=\frac12e^{-3\alpha}\Big\{} +e^{6\alpha}\tilde m^2\Big[3\varphi^2\Big(\tfrac12a_{\vec n,\epsilon}^2-2b_{\vec n,\epsilon}^2\Big)+6\varphi a_{\vec n,\epsilon}f_{\vec n,\epsilon}+f_{\vec n,\epsilon}^2\Big]\Big\}.
\end{align}
On the other hand, $\tilde H^{\vec n,\epsilon}_{|1}$ and $ \tilde H^{\vec n,\epsilon}_{\_1}$ are linear in the inhomogeneous perturbations. $\tilde H^{\vec n,\epsilon}_{|1}$ arises from the perturbation of the Hamiltonian constraint around the FRW model in full general relativity, while $ \tilde H^{\vec n,\epsilon}_{\_1}$ comes from the perturbation of the momentum (or diffeomorphisms) constraint. These linear perturbative constraints are given by
\begin{align}
\tilde{H}^{\vec n,\epsilon}_{|1} &= \frac{1}{2}\Big[-2\pi_\alpha\pi_{a_{\vec n,\epsilon}}+2\pi_\varphi\pi_{f_{\vec n,\epsilon}}-\big(\pi_\alpha^2+3\pi_\varphi^2\big)a_{\vec n,\epsilon}-\tfrac23\omega_n^2e^{4\alpha}(a_{\vec n,\epsilon}+b_{\vec n,\epsilon}) \nonumber\\
&\phantom{= \frac12e^{-3\alpha}\Big[} +e^{6\alpha}\tilde m^2\varphi(3\varphi a_{\vec n,\epsilon}+2f_{\vec n,\epsilon})\Big], \label{eq:H^n_|1}\\
\tilde{H}^{\vec n,\epsilon}_{\_1} &= \frac{1}{3}\big[-\pi_{a_{\vec n,\epsilon}}+\pi_{b_{\vec n,\epsilon}}+\pi_\alpha(a_{\vec n,\epsilon}+4b_{\vec n,\epsilon})+3\pi_\varphi f_{\vec n,\epsilon}\big]. \label{eq:H^n__1}
\end{align}
We note that $g_{\vec n,\epsilon}$ and $k_{\vec n,\epsilon}$ do not parameterize physical degrees of freedom, but they are instead the Lagrange multipliers associated with these linear perturbative constraints. Finally, it is worth remarking that, at the order of truncation adopted in the action, the perturbed system is symplectic, with canonical variables given by the zero-modes $\alpha$ and $\varphi$, the Fourier coefficients $\{X^{\vec n,\epsilon}_{q_l}\}\equiv\{a_{\vec n,\epsilon}, b_{\vec n,\epsilon}, f_{\vec n,\epsilon}\}$ (with $l=1,2,3$), and their corresponding momenta \cite{HH,hybridFRWflat}.

\section{Perturbations in terms of gauge-invariant variables}
\label{sec:inv}

As explained in the introduction, our first goal is to describe our perturbative system in terms of gauge-invariant variables without fixing any gauge freedom, preserving covariance at the level of the perturbative description. With this aim, we first focus our attention on the inhomogeneous sector of the phase space which contains the degrees of freedom of the perturbations. In the previous section, these degrees of freedom were parameterized by the variables $\{X^{\vec n,\epsilon}_{q_l}\}$, together with their canonically conjugate momenta $\{X^{\vec n,\epsilon}_{p_l}\}\equiv\{\pi_{a_{\vec n,\epsilon}}, \pi_{b_{\vec n,\epsilon}}, \pi_{f_{\vec n,\epsilon}}\}$, with $l$ running from 1 to 3. We will introduce a canonical transformation to describe these perturbations in terms of the MS variables, two suitable combinations of the linear perturbative constraints, and appropriate conjugate pairs of all of them. Since the intention is to respect the canonical structure of the set of elementary variables used for the perturbations, it is clear that we need to find, in particular, combinations of the linear perturbative constraints which commute, therefore abelianizing the perturbative constraint algebra. Once this abelianization is introduced, the fact that the MS variables are gauge invariant, and hence commute with the perturbative constraints, makes straightforward to find a complete set of compatible elementary variables for the inhomogeneous sector. The wanted canonical transformation is then attained with a convenient choice of conjugate variables. Later on, in Sec. \ref{sec:complete}, we will complete this transformation into a canonical one in our entire system, that is, considering not only the inhomogeneities but including also the homogeneous sector of the phase space, parameterized by the canonical variables for the zero-modes $\{\alpha,\pi_\alpha, \varphi,\pi_\varphi\}$. In total, the resulting Hamiltonian will be a linear combination of constraints, that are not only first-class (as usual), but furthermore that form an abelian algebra.

So, since in this section we are only interested in the canonical transformations of the inhomogeneous sector of our system and in the symplectic structure induced on it, we will consider momentarily that the homogeneous sector is a fixed background. With this  purpose, and only in that sense, we can ignore for the moment the Poisson brackets of the zero-mode variables both between themselves and with the perturbations. Hence, in this section we will operate with a Poisson bracket in the inhomogeneous sector defined accordingly as 
\begin{align}
 \{F,G\}_{(P)}\equiv\sum_{l,\vec n,\epsilon} \left(\frac{\partial F}{\partial X^{\vec n,\epsilon}_{q_l}}\frac{\partial G}{\partial X^{\vec n,\epsilon}_{p_l}}- \frac{\partial F}{\partial X^{\vec n,\epsilon}_{p_l}}\frac{\partial G}{\partial X^{\vec n,\epsilon}_{q_l}}\right),
\end{align}
where $F$ and $G$ are functions of the perturbations, and might also depend on the background variables.

\subsection{Canonical transformation for the perturbations}

Let us start by introducing the MS variables. In terms of the configuration modes $X^{\vec n,\epsilon}_{q_l}$, the modes of the MS field are given by \cite{mukhanov,hybridFRWflat,hybridFRWflatMS}
\begin{equation}\label{Mmode}
v_{\vec n,\epsilon} = e^\alpha\left[f_{\vec n,\epsilon}+\frac{\pi_{\varphi}}{\pi_\alpha}(a_{\vec n,\epsilon}+b_{\vec n,\epsilon})\right].
\end{equation}
It is straightforward to check that these modes indeed are gauge invariant, as they Poisson commute with the linear perturbative constraints: 
\begin{equation}
\{v_{\vec n,\epsilon},  \tilde H^{\vec n',\epsilon'}_{|1}\}_{(P)}=0=\{v_{\vec n,\epsilon},  \tilde{H}^{\vec n',\epsilon'}_{\_1}\}_{(P)}.
\end{equation}

We would like to complete these modes to a set of compatible elementary variables for the description of the perturbations. Since the MS variables are gauge invariant, it is natural to try this completion by considering combinations of the linear perturbative constraints, with which they commute. Besides, in this way, the information about the perturbative constraints will be straightforwardly incorporated in our system in terms of elementary variables. In particular, imposing them quantum mechanically will be a direct task. The fact that we want to extract two compatible variables from the perturbative constraints means, as we have already pointed out, that we have to abelianize the algebra of those constraints (and a posteriori, the entire constraint algebra of the perturbed FRW system). With this aim in mind, we notice that the only non-vanishing Poisson brackets between them is $\{\tilde H^{\vec n,\epsilon}_{\_1}, \tilde H^{\vec n,\epsilon}_{|1}\}_{(P)}=e^{3\alpha}  H_{|0}$. It is then easy to abelianize their algebra: We only need to replace $\tilde H^{\vec n,\epsilon}_{|1}$ with the combination
\begin{align}
\breve{H}^{\vec n,\epsilon}_{|1}&=\tilde H^{\vec n,\epsilon}_{|1}-3e^{3\alpha}  H_{|0} a_{\vec n,\epsilon}\nonumber\\
&=-\pi_\alpha\pi_{a_{\vec n,\epsilon}}+\pi_\varphi\pi_{f_{\vec n,\epsilon}}+\big(\pi_\alpha^2-3\pi_\varphi^2-\tfrac13\omega_n^2e^{4\alpha}\big)a_{\vec n,\epsilon}-\tfrac13\omega_n^2e^{4\alpha}b_{\vec n,\epsilon} +e^{6\alpha}\tilde m^2\varphi f_{\vec n,\epsilon}.
\end{align}
Actually, this change amounts to a redefinition of the zero-mode of the lapse function. Indeed, in the action of the system and up to the quadratic order in perturbations that we are working with, we can rewrite the Hamiltonian \eqref{eq:Hamiltonian} as 
\begin{equation}\label{eq:Hamiltonian2}
H =\breve N_0\Big[ H_{|0}+\sum_{\vec n,\epsilon} H^{\vec n,\epsilon}_{|2}\Big]+\sum_{\vec n,\epsilon} g_{\vec n,\epsilon}\breve{H}^{\vec n,\epsilon}_{|1}+\sum_{\vec n,\epsilon}k_{\vec n,\epsilon}\tilde 
H^{\vec n,\epsilon}_{\_1},
\end{equation}
with the new zero-mode of the lapse function acquiring contributions (of quadratic order) from the inhomogeneities:
\begin{equation}\label{N0re}
 \breve N_0 =N_0+3 e^{3\alpha}\sum_{\vec n,\epsilon}g_{\vec n,\epsilon}a_{\vec n,\epsilon}.
\end{equation}
Note that the MS variables $v_{\vec n,\epsilon}$ remain invariant with respect to the new set of constraints. In summary, the set of variables $\{v_{\vec n,\epsilon},\breve H^{\vec n,\epsilon}_{|1},\tilde{H}^{\vec n,\epsilon}_{\_1}\}$ provides a  parameterization of the inhomogeneous configuration space (inasmuch as that all the variables are compatible) in terms of constraints and gauge invariants, as we wanted.

We now complete the canonical transformation in the inhomogeneous sector by determining variables canonically conjugate to our new set. It is straightforward to check that 
\begin{align}
C^{\vec n,\epsilon}_{\_1}=3 b_{\vec n,\epsilon} 
\end{align}
is canonically conjugate to $\tilde H^{\vec n,\epsilon}_{\_1}$, namely $\{C^{\vec n,\epsilon}_{\_1},\tilde H^{\vec n,\epsilon}_{\_1}\}_{(P)}=1$, while it Poisson commutes with $v_{\vec n',\epsilon'}$ and $\breve{H}^{\vec n',\epsilon'}_{|1}$. Finding the other canonical variables requires a bit more of work. We skip the details of the calculation and encourage the reader to check that one can choose
\begin{align}
C^{\vec n,\epsilon}_{|1}=-\frac1{\pi_\alpha}(a_{\vec n,\epsilon}+ b_{\vec n,\epsilon}),
\end{align}
as the variable conjugate to the constraint $\breve{H}^{\vec n,\epsilon}_{|1}$, in the sense that  $\{C^{\vec n,\epsilon}_{|1},\breve{H}^{\vec n,\epsilon}_{|1}\}_{(P)}=1$, whereas 
\begin{align}
 \pi_{v_{\vec n,\epsilon}}=e^{-\alpha}\bigg[\pi_{f_{\vec n,\epsilon}}+\frac1{\pi_\varphi}\Big(e^{6\alpha}\tilde{m}^2 \varphi f_{\vec n,\epsilon}+3\pi_\varphi^2  b_{\vec n,\epsilon} \Big)\bigg]
\end{align}
is a momentum conjugate to the MS variable $v_{\vec n,\epsilon}$, that is $\{v_{\vec n,\epsilon}, \pi_{v_{\vec n,\epsilon}} \}_{(P)}=1$.

For convenience, we will assign the role of configuration variables to $C^{\vec n,\epsilon}_{\_1}$ and $C^{\vec n,\epsilon}_{|1}$, and view the constraints $\tilde H^{\vec n,\epsilon}_{\_1}$ and $\breve H^{\vec n,\epsilon}_{|1}$ as momenta (this will simplify the discussion about the imposition of the perturbative constraints à la Dirac in the quantization of the system). Furthermore, we will use a compact notation for the new set of canonical variables for the perturbations, namely we define 
\begin{eqnarray}
\{V^{\vec n,\epsilon}_{q_l}\}&\equiv&\{v_{\vec n,\epsilon}, C^{\vec n,\epsilon}_{|1},C^{\vec n,\epsilon}_{\_1}\},\\
\{V^{\vec n,\epsilon}_{p_l}\}&\equiv&\{\pi_{v_{\vec n,\epsilon}}, \breve H^{\vec n,\epsilon}_{|1}, \tilde H^{\vec n,\epsilon}_{\_1}\}.
\end{eqnarray} 
In this way, our previous configuration variables $X^{\vec n,\epsilon}_{q_l}$ and the new ones $V^{\vec n,\epsilon}_{q_l}$ are related by means of a contact transformation.

\subsection{Redefinition of the MS momenta}
\label{MSmomentum}

In Sec. \ref{sec:hybrid} we will carry out a Fock quantization of the perturbations, and in particular of the MS gauge-invariant field. 
If one reduces the system classically so that it gets described by QFT in a curved background, the resulting MS field verifies the Klein-Gordon equation of a massive scalar field (with time-dependent mass) propagating in an ultrastatic spacetime with compact spatial topology.
Remarkably, a series of studies on the uniqueness of the Fock representation for Klein-Gordon fields of this type \cite{uniqueness1,uniqueness2,uniquenessother} proves that the use of the MS field to describe the perturbations allows for a unitary implementation of the quantum dynamics of the field, and that any other parameterization for the perturbations that includes a time-dependent rescaling of the field (as the one employed in Ref. \cite{dressed}), prevents the dynamics for being unitarily implementable in the quantum theory \cite{uniqueperturb}. The unitary implementation of the dynamics is possible in a class of (unitarily) equivalent Fock representations with vacua that are invariant under the isometries of the three-torus. Moreover, these results require a specific choice of momentum for the MS variable among all the canonical pairs, namely the evolution can be implemented unitarily as long as the MS modes verify the equation of motion 
\begin{align}\label{eqhamipi}
 \dot v_{\vec n,\epsilon}=\pi_{v_{\vec n,\epsilon}}
\end{align}
at the considered perturbative order. Here the dot denotes, as usual, derivative with respect to the time coordinate $t$. Since the evolution equations are generated by the Hamiltonian of the system and the MS variables commute with the perturbative constraints, it is not difficult to convince oneself that the above condition on the choice of momenta is achieved only if the zero-mode of the Hamiltonian constraint, that contains quadratic contributions of the perturbations, does not include any linear term in the momenta conjugate to the MS variables.\footnote{Actually, the condition happens to be not only necessary, but also sufficient.} 

We want to adhere to the above parameterization of the MS field, and therefore we want to choose its conjugate variable so as to eliminate all terms that are linear in the MS field momentum from the Hamiltonian constraint. With that aim, we need to modify the momentum mode $\pi_{v_{\vec n,\epsilon}}$ by taking advantage of the freedom in adding to it a linear contribution of the MS configuration variable $v_{\vec n,\epsilon}$. Thus, we introduce a change of the form
\begin{align}
 \pi_{v_{\vec n,\epsilon}}\; \rightarrow \; \breve \pi_{v_{\vec n,\epsilon}}=\pi_{v_{\vec n,\epsilon}}+ Fv_{\vec n,\epsilon},
\end{align}
where $F$ is a function of the homogeneous variables $(\alpha,\pi_\alpha, \varphi,\pi_\varphi)$. We now have to determine this function by appealing to condition \eqref{eqhamipi}.
Notice that, by construction, the new set $\{\breve V^{\vec n,\epsilon}_{p_l}\}\equiv\{\breve\pi_{v_{\vec n,\epsilon}}, \breve H^{\vec n,\epsilon}_{|1}, \tilde H^{\vec n,\epsilon}_{\_1}\}$ still provides a canonical set together with  $\{V^{\vec n,\epsilon}_{q_l}\}$.

In order to find the explicit expression of $F$ we can proceed in a simple way as follows. For a moment, we go to the longitudinal gauge, which is picked out by the pair of conditions $ b_{\vec n,\epsilon}=0$ and $\pi_{a_{\vec n,\epsilon}}=\pi_\alpha a_{\vec n,\epsilon}+3\pi_\varphi f_{\vec n,\epsilon}$. In this gauge, we match the resulting MS momentum with the result that was obtained (precisely in that gauge) in Ref. \cite{hybridFRWflatMS}. This procedure turns out to provide a unique and well determined answer, showing the consistency of our calculations. Moreover, the function $F$ is actually mode independent (therefore our notation for it). Its form is
\begin{align}
F=-e^{-2\alpha}\bigg(\frac1{\pi_\varphi} e^{6\alpha}\tilde{m}^2 \varphi+\pi_\alpha+3\frac{\pi_\varphi^2}{\pi_\alpha} \bigg).
\end{align}
Thus, the modes of the new MS momentum are
\begin{eqnarray}
\breve \pi_{v_{\vec n,\epsilon}}&=&e^{-\alpha}\bigg[\pi_{f_{\vec n,\epsilon}}+\frac1{\pi_\varphi}\Big(e^{6\alpha}\tilde{m}^2 \varphi f_{\vec n,\epsilon}+3\pi_\varphi^2  b_{\vec n,\epsilon} \Big)\bigg] \nonumber\\
&-&e^{-2\alpha}\bigg(\frac1{\pi_\varphi} e^{6\alpha}\tilde{m}^2 \varphi+\pi_\alpha+3\frac{\pi_\varphi^2}{\pi_\alpha} \bigg)v_{\vec n,\epsilon}.
\end{eqnarray}
In order to simplify the notation, in the following we will use again the symbol $\pi_{v_{\vec n,\epsilon}}$ to denote these redefined momentum modes and $ V^{\vec n,\epsilon}_{p_l}$ for the corresponding set of new momenta.

\subsection{Inversion of the canonical transformation}

For completeness, we finish this section by giving explicitly the expression of the original perturbative variables  $\{X^{\vec n,\epsilon}_{l}\}\equiv \{X^{\vec n,\epsilon}_{q_l}, X^{\vec n,\epsilon}_{p_l}\}$ in terms of the new ones $\{V^{\vec n,\epsilon}_{l}\}\equiv \{V^{\vec n,\epsilon}_{q_l}, V^{\vec n,\epsilon}_{p_l}\}$. The result is
\begin{subequations}\label{eq:transf}
\begin{align}
a_{\vec n,\epsilon}&=-\pi_\alpha V^{\vec n,\epsilon}_{q_2}-\frac13V^{\vec n,\epsilon}_{q_3},\label{a}\\
b_{\vec n,\epsilon}&=\frac13V^{\vec n,\epsilon}_{q_3},\\
f_{\vec n,\epsilon}&=e^{-\alpha} V^{\vec n,\epsilon}_{q_1}+\pi_\varphi V^{\vec n,\epsilon}_{q_2},\\
\pi_{a_{\vec n,\epsilon}}&=-\frac1{\pi_\alpha} V^{\vec n,\epsilon}_{p_2}+\frac{\pi_\varphi}{\pi_\alpha} e^\alpha  V^{\vec n,\epsilon}_{p_1}+\frac{e^{-\alpha} }{\pi_\alpha}\bigg( e^{6\alpha} \tilde{m}^2 \varphi +\pi_\varphi \pi_\alpha+ 3 \frac{\pi_\varphi^3}{\pi_\alpha} \bigg) V^{\vec n,\epsilon}_{q_1}\nonumber\\
&+\bigg(3\pi_\varphi^2+\frac13 e^{4\alpha} \omega_n^2 -\pi_\alpha^2\bigg) V^{\vec n,\epsilon}_{q_2}-\frac{1}{3} \pi_\alpha V^{\vec n,\epsilon}_{q_3},\\
\pi_{b_{\vec n,\epsilon}}&=3V^{\vec n,\epsilon}_{p_3}-\frac1{\pi_\alpha} V^{\vec n,\epsilon}_{p_2}+\frac{\pi_\varphi}{\pi_\alpha} e^\alpha  V^{\vec n,\epsilon}_{p_1}+\frac{e^{-\alpha} }{\pi_\alpha}\bigg( e^{6\alpha} \tilde{m}^2 \varphi -2\pi_\varphi \pi_\alpha+ 3 \frac{\pi_\varphi^3}{\pi_\alpha} \bigg) V^{\vec n,\epsilon}_{q_1}\nonumber\\
&+\frac13 e^{4\alpha} \omega_n^2 V^{\vec n,\epsilon}_{q_2}-\frac{4}{3} \pi_\alpha V^{\vec n,\epsilon}_{q_3},\\
\pi_{f_{\vec n,\epsilon}}&=e^{\alpha} V^{\vec n,\epsilon}_{p_1}+e^{-\alpha}\bigg( \pi_\alpha+3 \frac{\pi_\varphi^2}{\pi_\alpha} \bigg) V^{\vec n,\epsilon}_{q_1}-e^{6\alpha} \tilde{m}^2 \varphi V^{\vec n,\epsilon}_{q_2}-\pi_\varphi V^{\vec n,\epsilon}_{q_3}.
\end{align}
\end{subequations}

\section{Full canonical set in terms of gauge-invariant variables}
\label{sec:complete}

Recall that in the previous section we regarded the zero-modes, described by the variables of the unperturbed FRW model $\{w^a_q\}\equiv \{\alpha, \varphi\}$ and  $\{w^a_p\}\equiv \{\pi_\alpha,\pi_\varphi\}$ ($a=1,2$), as corresponding to a fixed background. Now, we proceed to complete our canonical transformation in the entire system, including these zero-mode variables. In this way we will succeed in describing our perturbed system in terms of gauge invariants, without any gauge fixing, while keeping its full canonical structure. 

\subsection{Canonical transformation of the zero-modes}

The action of our system (truncated at quadratic order in perturbations) in terms of the original parameterization of the inhomogeneous sector has the form
\begin{align}
 S=\int \text{d}t \bigg[\sum_{a} \dot w^a_q w^a_p+\sum_{l,\vec n,\epsilon} \dot X^{\vec n,\epsilon}_{q_l}  X^{\vec n,\epsilon}_{p_l} + H(w^a, X^{\vec n,\epsilon}_l)\bigg],
\end{align}
possibly up to surface terms which are not relevant for our discussion. Here, $H(w^a, X^{\vec n,\epsilon}_l)$ is the total Hamiltonian expressed in terms of the original variables employed in Sec. \ref{sec:system}, $\{w^a\}\equiv \{w^a_q,w^a_p\}$ and $\{X^{\vec n,\epsilon}_{l}\}\equiv \{X^{\vec n,\epsilon}_{q_l}, X^{\vec n,\epsilon}_{p_l}\}$, as in Eq. \eqref{eq:Hamiltonian2}.
Let us focus on the Legendre term that contains the information about the symplectic structure of the full system, that we call $W$:
\begin{align}\label{eq:W1}
 W=\int \text{d}t \bigg[ \sum_{a} \dot w^a_q w^a_p+\sum_{l,\vec n,\epsilon} \dot X^{\vec n,\epsilon}_{q_l}  X^{\vec n,\epsilon}_{p_l}\bigg].
\end{align}
Our goal is to find a canonical transformation relating the previous variables with new zero-modes $\tilde w^a\equiv (\tilde w^a_q,\tilde w^a_p)$ such that $W$ retains its canonical form when expressed in terms of the gauge invariants for the perturbations, namely
\begin{align}\label{eq:W2}
 W=\int \text{d}t \bigg[ \sum_{a} \dot {\tilde w}^a_q \tilde w^a_p+\sum_{l,\vec n,\epsilon} \dot V^{\vec n,\epsilon}_{q_l}  V^{\vec n,\epsilon}_{p_l} \bigg].
\end{align}

Instead of reproducing all the details of the lengthy calculation that allows one to deduce the form of the desired transformation (and which is essentially based on the consideration of our change of perturbative variables as a canonical transformation in the inhomogeneous sector that depends on a series of time-dependent {\it{external}} variables, describing the homogeneous degrees of freedom), we will only sketch the main steps in the derivation. First, we notice that the relations $ X^{\vec n,\epsilon}_{l}= X^{\vec n,\epsilon}_{l}( V^{\vec n,\epsilon}_{m})$ given in Eqs. \eqref{eq:transf} do not mix different modes and are linear. Therefore, the original variables $X^{\vec n,\epsilon}_{l}$ can be expressed in the following way
\begin{align}\label{eq:chain}
 X^{\vec n,\epsilon}_{l}=\sum_m \bigg(\frac{\partial X^{\vec n,\epsilon}_{l}}{\partial V^{\vec n,\epsilon}_{q_m}}V^{\vec n,\epsilon}_{q_m}+\frac{\partial X^{\vec n,\epsilon}_{l}}{\partial V^{\vec n,\epsilon}_{p_m}}V^{\vec n,\epsilon}_{p_m}\bigg),
\end{align}
where the partial derivatives are functions of the zero-modes $w^a$ and the frequency $\omega_n$ only  (and $m=1,2,3$). Taking this relation into account in the derivatives with respect to time that appear in $W$, performing several time integrations by parts, disregarding total time derivatives which contribute at most by surface terms at initial and final times (assuming a well posed variational principle for the system), and truncating up to quadratic order in perturbations, as well as recalling that $\{V^{\vec n,\epsilon}_{l}\}$ is a canonical set as long as one ignores the homogeneous sector, one can obtain the expression of the new canonical homogeneous variables $\tilde w ^a$ as functions of the old ones $ w ^a$ and of the new perturbative variables $V^{\vec n,\epsilon}_{l}$ (or of $X^{\vec n,\epsilon}_{l}$ if preferred). In this way, one gets new zero-mode variables which are given by the original ones plus some additional contributions that are quadratic in the perturbations. This result can be inverted at the considered perturbative order, allowing one to express the original zero-mode variables as functions of the new phase-space variables for the entire system (zero-modes plus inhomogeneities). With this procedure, one arrives at
\begin{subequations}\label{eq:transf-background}
\begin{align}\label{homochange1}
w^a_q&= \tilde w^a_q - \frac12 \sum_{l,\vec n,\epsilon} \bigg[X^{\vec n,\epsilon}_{q_l} \frac{\partial X^{\vec n,\epsilon}_{p_l}}{\partial_{\tilde w^a_p}}- \frac{\partial  X^{\vec n,\epsilon}_{q_l}}{\partial_{\tilde w^a_p}} X^{\vec n,\epsilon}_{p_l}\bigg],\\ \label{homochange2}
 w^a_p&= \tilde w^a_p + \frac12 \sum_{l,\vec n,\epsilon} \bigg[X^{\vec n,\epsilon}_{q_l} \frac{\partial X^{\vec n,\epsilon}_{p_l}}{ \partial_{\tilde w^a_q}}- \frac{\partial X^{\vec n,\epsilon}_{q_l}}{\partial_{\tilde w^a_q}}X^{\vec n,\epsilon}_{p_l}\bigg].
\end{align}
\end{subequations}
In these expressions we have to understand the original variables $X^{\vec n,\epsilon}_{l}$ for the perturbations as functions of the new ones $V^{\vec n,\epsilon}_{l}$, as given by Eqs. \eqref{eq:transf}, with $\{w^a\}=\{\alpha,\pi_\alpha, \varphi,\pi_\varphi\}$ replaced with $\{\tilde w^a\}=\{\tilde\alpha,\tilde\pi_\alpha, \tilde\varphi,\tilde\pi_\varphi\}$ in those equations. Note that this replacement is consistent within our truncations, since old and new zero-modes differ in terms that are quadratic in perturbations. In this sense, let us also clarify that the partial derivatives in the above identities must be taken keeping  $V^{\vec n,\epsilon}_{l}$ constant. 
In Appendix \ref{app} we give the explicit expressions of the zero-modes $w^a$ in terms of the new phase-space variables $\tilde w^a$ and $V^{\vec n,\epsilon}_{l}$.

\subsection{Hamiltonian in terms of gauge-invariant variables}

Once we know the canonical transformation relating the original phase-space parameterization, $\{w^a, X^{\vec n,\epsilon}_{l}\}$, with the new one that uses gauge invariants for the perturbations, $\{\tilde w^a, V^{\vec n,\epsilon}_{l}\}$, we can obtain the form of the Hamiltonian in the new formulation. In particular, let us consider the zero-mode of the Hamiltonian constraint. In the original variables, this zero-mode is given in Eq. \eqref{eq:Hamiltonian2} by the sum 
\begin{equation}
H_{|0}(w^a)+ \sum_{\vec n,\epsilon} H^{\vec n,\epsilon}_{|2}  \big(w^a, X^{\vec n,\epsilon}_{l}\big),
\end{equation} 
which contains the contribution of the homogeneous FRW model and the correction quadratic in perturbations. 

Since the difference between the original and new variables for the homogeneous sector is precisely of quadratic order in the perturbations, the substitution of the relations between the old and new variables for our system leads to the following new expression for the zero-mode of the Hamiltonian constraint:
\begin{equation}\label{newzeroham}
H_{|0}(\tilde{w}^a)+ \sum_b \big(w^b-\tilde{w}^b\big) \frac{\partial H_{|0}}{\partial \tilde{w}^b}(\tilde{w}^a) + \sum_{\vec n,\epsilon} H^{\vec n,\epsilon}_{|2}  \big[\tilde{w}^a, X^{\vec n,\epsilon}_{l}(\tilde{w}^a, V^{\vec n,\epsilon}_{l}) \big],
\end{equation} 
where the difference $w^a-\tilde{w}^a$ must be regarded as a function of the new phase-space variables, given by the last terms in the two expressions \eqref{homochange1} and \eqref{homochange2}. Note that this difference is a sum of independent contributions of each of the inhomogeneous modes, so that we can write 
\begin{equation}
w^a-\tilde{w}^a \equiv \sum_{\vec n,\epsilon} \Delta \tilde{w}^a_{\vec n,\epsilon}.
\end{equation} 
Besides, evaluation of $H_{|0}$ and $H^{\vec n,\epsilon}_{|2} $ at the new zero-mode variables $\tilde{w}^a$ is done in Eq. \eqref{newzeroham} by simply replacing the original phase-space coordinates $w^a$ directly with these new ones in the functional dependence of the considered contributions to the constraint. To derive the above expression, it suffices to expand $H_{|0}$ in series around the new zero-mode variables and recall that we are truncating the total Hamiltonian at quadratic order in the perturbations. From these considerations, we see that the quadratic contribution of the perturbations to the zero-mode of the Hamiltonian constraint in our new variables takes the form $\sum_{\vec n,\epsilon} \tilde{H}^{\vec n,\epsilon}_{|2}$, with  
\begin{equation}
\tilde{H}^{\vec n,\epsilon}_{|2} = H^{\vec n,\epsilon}_{|2} + \sum_a \Delta \tilde{w}^a_{\vec n,\epsilon}\frac{ \partial H_{|0} }{ \partial \tilde{w}^a }  .  
\end{equation}
Indeed, this is the form that one would expect if one regarded the change of variables for the perturbations as a canonical transformation that depends on a series of time-dependent {\it{external}} variables (namely, the zero-modes) with dynamics governed by the  Hamiltonian $H_{|0}$ \cite{Langlois}.

If one carries out the calculation explicitly, one gets the following quadratic contribution to the constraint of our system:
\begin{align}\label{eq:newH2}
\tilde{H}^{\vec n,\epsilon}_{|2}=\breve H^{\vec n,\epsilon}_{|2} +  F_{|2}^{\vec n,\epsilon} H_{|0}+F_{|1}^{\vec n,\epsilon} V^{\vec n,\epsilon}_{p_2}+\bigg(F_{\_1}^{\vec n,\epsilon}-3 \frac{e^{-3{\tilde\alpha}}}{\pi_{\tilde\alpha}} V^{\vec n,\epsilon}_{p_2} + \frac92e^{-3{\tilde\alpha}} V^{\vec n,\epsilon}_{p_3}\bigg)V^{\vec n,\epsilon}_{p_3}.
\end{align}
In this expression we have defined 
\begin{subequations}
\begin{align}\label{eq:H2MS}
\breve H^{\vec n,\epsilon}_{|2}&=\frac{e^{-{\tilde\alpha}}}{2}\bigg[ \omega_n^2 + e^{-4{\tilde\alpha}}\pi_{\tilde\alpha}^2+\tilde{m}^2 e^{2{\tilde\alpha}}\left(1+15{\tilde\varphi}^2-12{\tilde\varphi} \frac{\pi_{\tilde\varphi}}{\pi_{\tilde\alpha}}-18 e^{6{\tilde\alpha}}\tilde{m}^2 \frac{{\tilde\varphi}^4}{\pi_{\tilde\alpha}^2} \right)\bigg] (V^{\vec n,\epsilon}_{q_1})^2 \nonumber\\
&+\frac{e^{-{\tilde\alpha}}}{2} (V^{\vec n,\epsilon}_{p_1})^2,\\
F_{|2}^{\vec n,\epsilon}&= \frac32 e^{-2{\tilde\alpha}}\bigg( 1- 9 \frac{\pi_{\tilde\varphi}^2}{\pi_{\tilde\alpha}^2}+12 e^{6{\tilde\alpha}}\tilde{m}^2 \frac{{\tilde\varphi}^2}{\pi_{\tilde\alpha}^2}\bigg) (V^{\vec n,\epsilon}_{q_1})^2 + \frac12 e^{-2{\tilde\alpha}}\left( e^{4{\tilde\alpha}}\omega_n^2 -9\pi_{\tilde\varphi}^2 \right)(V^{\vec n,\epsilon}_{q_2})^2\nonumber\\
&-3 \frac{e^{-2{\tilde\alpha}}}{\pi_{\tilde\alpha}^2}\bigg[\left(\pi_{\tilde\alpha}^2 +3\pi_{\tilde\varphi}^2\right)\pi_{\tilde\varphi}+e^{6{\tilde\alpha}}\tilde{m}^2 {\tilde\varphi} \pi_{\tilde\alpha}\bigg]V^{\vec n,\epsilon}_{q_1}V^{\vec n,\epsilon}_{q_2}-3\frac{\pi_{\tilde\varphi}}{\pi_{\tilde\alpha}}V^{\vec n,\epsilon}_{p_1}V^{\vec n,\epsilon}_{q_2},\\
F_{|1}^{\vec n,\epsilon}&=\frac12 \frac{e^{-4{\tilde\alpha}}}{\pi_{\tilde\alpha}}\bigg[6 \pi_{\tilde\varphi} V^{\vec n,\epsilon}_{q_1}+\left(6 e^{3{\tilde\alpha}} H_{|0} +5 \pi_{\tilde\alpha}^2\right)V^{\vec n,\epsilon}_{q_2}\bigg],\\
F_{\_1}^{\vec n,\epsilon}&=3 \frac{e^{-4{\tilde\alpha}}}{\pi_{\tilde\alpha}^2}\left( e^{6{\tilde\alpha}}\tilde{m}^2 {\tilde\varphi} \pi_{\tilde\alpha} +3\pi_{\tilde\varphi}^3 -2 \pi_{\tilde\alpha}^2\pi_{\tilde\varphi} \right)V^{\vec n,\epsilon}_{q_1}+ \omega_n^2 V^{\vec n,\epsilon}_{q_2}-\frac32 e^{-6{\tilde\alpha}}\pi_{\tilde\alpha}  V^{\vec n,\epsilon}_{q_3}\nonumber\\
&-3 e^{-2{\tilde\alpha}}\frac{\pi_{\tilde\varphi}}{\pi_{\tilde\alpha}}V^{\vec n,\epsilon}_{p_1}.
\end{align}
\end{subequations}
In the total Hamiltonian, we can account for the term $\sum_{\vec n, \epsilon} F_{|2}^{\vec n,\epsilon} H_{|0}$ at the considered perturbative order by means of a new redefinition of the zero-mode of the lapse function, with the new one given by $\bar N_0=\breve N_0 (1 +\sum_{\vec n,\epsilon} F_{|2}^{\vec n,\epsilon})$ [see Eq. \eqref{N0re}]. The other terms in Eq. \eqref{eq:newH2} different from $\breve H^{\vec n,\epsilon}_{|2}$ contribute to the perturbative linear constraints and their presence amounts to a redefinition of the corresponding Lagrange multipliers, that we now denote as $G_{\vec n,\epsilon}$ and $K_{\vec n,\epsilon}$. In summary, we obtain at our quadratic order a total Hamiltonian constraint of the form
\begin{equation}\label{eq:Hamiltonian3}
H =\bar N_0\Big[ H_{|0}+\sum_{\vec n,\epsilon} \breve H^{\vec n,\epsilon}_{|2}\Big]+\sum_{\vec n,\epsilon} G_{\vec n,\epsilon} V^{\vec n,\epsilon}_{p_2}+ \sum_{\vec n,\epsilon} K_{\vec n,\epsilon} V^{\vec n,\epsilon}_{p_3} .
\end{equation}
In Appendix \ref{app} we give the explicit expressions of the original lapse function $N$ and shift co-vector $N_i$  in terms of the new phase-space variables $\{\tilde w^a, V^{\vec n,\epsilon}_{l}\}$ and Lagrange multipliers $\{\bar N_0,  G_{\vec n,\epsilon},  K_{\vec n,\epsilon}\}$.

As we see, the terms $\breve H^{\vec n,\epsilon}_{|2}$ provide finally the contributions quadratic in perturbations to the zero-mode of the Hamiltonian constraint in the constructed gauge-invariant formulation. As expected, these terms are precisely those that depend exclusively on the MS variables and their momenta, and hence they are gauge-invariant quantities. For obvious reasons, we call their sum the MS Hamiltonian. Moreover, as we had anticipated, these terms contain no linear contribution of the momenta of the MS variables. In fact, the only contribution from these momenta is quadratic, and its mode-independent  coefficient is constant up to a power of the scale factor. We notice as well that the expression given for $\breve H^{\vec n,\epsilon}_{|2}$ is linear in the momentum of the zero-mode of the scalar field, $\pi_{\tilde\varphi}$. This linear expression has been obtained by taking advantage of the identity $\pi_{\varphi}^2=2 H_{|0} e^{3{\alpha}} +\pi_{\alpha}^2-e^{6{\alpha}}\tilde{m}^2 {\varphi}^2$, that can be used in Eq. \eqref{eq:newH2} at the coast of redefining the zero-mode of the lapse function at the considered order in perturbation theory. Thanks to this, in Sec. \ref{sec:hybrid} we will be able to interpret the zero-mode of the Hamiltonian constraint in a certain approximation as a Schr\"odinger-like equation that dictates the quantum evolution of the inhomogeneities in some family of states. 

\section{Hybrid quantization and Born-Oppenheimer ansatz}
\label{sec:hybrid}

We can now proceed to quantize the symplectic manifold which describes our cosmological system, and to impose the classical constraints à la Dirac, i.e., as operators that annihilate physical states in the quantum theory.
We recall that, in our classical formulation, we have split the phase space in homogeneous and inhomogeneous sectors, starting with the expansion in modes given by the eigenfunctions of the Laplace-Beltrami operator on the spatial sections. The homogeneous sector describes the zero-modes, and its degrees of freedom can be parameterized by the canonical variables $\{\tilde{w}^a\}=\{\tilde\alpha,\pi_{\tilde\alpha},\tilde\varphi,\pi_{\tilde\varphi}\}$. On the other hand, the inhomogeneous sector describes all the non-zero modes of the perturbations, and can be parameterized by $\{V^{\vec n,\epsilon}_l\}=\{ V^{\vec n,\epsilon}_{q_l}, V^{\vec n,\epsilon}_{p_l}\}$, that is a set specially selected to make manifest the covariance of the formulation at the considered perturbative level. In this way we have prepared the classical system to now carry out a (generalized) hybrid quantization, in which one adopts specific quantizations for both the homogeneous and inhomogeneous parts. On the tensor product of the corresponding representation spaces, one imposes the constraints quantum mechanically, so that the total system is not at all trivial. 

In fact, it seems natural to assume that there exists some regime of the quantum dynamics, in between a fully quantum gravity regime and the regime where QFT in fixed curved backgrounds makes sense, in which the most relevant quantum effects of the geometry are those affecting the zero-modes, and that perturbations, while still having a quantum description, can be represented in more standard ways, like e.g. as in QFT in curved spacetimes. Hence, we can adhere to a hybrid quantization based on this hypothesis, adopting a Fock quantization for the MS gauge invariants (and possibly for the rest of variables for the perturbations) and, in principle, a different type of quantization, of quantum gravity nature, for the homogeneous sector. Note that this hybrid approach is different from treating the perturbations as a test field propagating in a quantum background, which is the viewpoint adopted in the {\it{dressed metric}} proposal in Ref. \cite{dressed}. Rather, the inhomogeneities and 
 the homogeneous sector, even being represented with different quantization methods, form a symplectic manifold that is quantized as a whole. The Hamiltonian constraint of the system affects both sectors, encoding the backreaction of the perturbations on the homogeneous background inasmuch as this is kept in the quadratic truncation of the action. 

In this section we will maintain a general discussion without specifying the concrete quantization of the homogeneous sector. Moreover, most of the details are easily extensible to other quantizations of our MS variables different from the Fock (or QFT-like) quantization that we adopt here. We will simply assume that the quantization of the zero-mode variables provides a representation for the canonical commutation relations such that the operators for the homogeneous FRW geometry commute with those representing the homogeneous sector of the matter field, and in turn all of them commute with the elementary operators representing the variables for the inhomogeneities. We will denote $\mathcal{H}_\text{kin}^\text{grav}\otimes \mathcal{H}_\text{kin}^\text{matt}$ the kinematical Hilbert space for the homogeneous sector, such that the operators representing the homogeneous FRW geometry are defined on $\mathcal{H}_\text{kin}^\text{grav}$, while the operators for the homogeneous matter sector are defined on $\mathcal{H}_\text{kin}^\text{matt}$. Finally, concerning the quantization of the homogeneous matter sector, we will also assume for simplicity that the operators that represent functions of $\tilde\varphi$ act by mere multiplication, while the operator representing $\pi_{\tilde\varphi}$ acts as a generalized derivative.

In previous applications of this strategy of a hybrid quantization \cite{hybridFRW,hybridFRWflat,hybridFRWflatMS,hybridgowdy}, the polymeric representation of LQC was adopted for the background geometry, together with a standard Schr\"odinger representation for the zero-mode $\tilde\varphi$ of the matter field. Although we will briefly explain the particularization of our study to such a representation in the next section, we intend to make here a much more general discussion, by no means restricted to the framework of LQC. Therefore, our analysis generalizes the quantum treatment presented in Ref. \cite{hybridFRWflatMS} in two important ways: a) instead of reducing the system classically, we carry the linear perturbative constraints to the quantum theory, with no gauge fixing; and b) we do not adopt here any specific representation for the homogeneous sector, and in particular for the FRW geometry. In this sense, the formulation can be adapted to any quantization proposal in FRW cosmology. The physics resulting from each specific proposal might be discussed and compared afterwards.

\subsection{Fock representation for the perturbations}

For the inhomogeneous modes describing the MS gauge invariants, we adopt here a Fock representation similar to that of Refs. \cite{hybridFRWflat,hybridFRWflatMS}. Remarkably, this 
quantization is selected by the criteria of: a) invariance of the vacuum under the spatial isometries, and b) unitary implementability of the dynamics in the regime in which one recovers a QFT in a curved background for the MS field (in any finite time interval) \cite{uniqueness1,uniqueness2}. As mentioned in Subsec. \ref{MSmomentum}, these criteria remove the ambiguity in a possible rescaling of the MS modes by functions of the homogeneous variables, and they assure a unitary implementability of the evolution and a standard quantum mechanical interpretation in the regions where a QFT in a (generally effective) background is recovered. Actually, these results require the spatial sections to be compact; that is why we choose our flat model to have three-torus topology. More concretely, the above 
criteria pick out not just a Fock representation, but a family of them such that they are all unitarily equivalent \cite{uniqueness2}. 
In particular, this family contains the representation in which the annihilation-like variables, $a_{\vec n,\epsilon}$, and the creation-like variables, $a^{\dagger}_{\vec n,\epsilon}$, are those naturally associated with harmonic oscillators of frequency $\omega_n$.
A more specific choice of Fock representation among the privileged family selected by those criteria can be done on the basis of further requirements not just on operators that are constructed from exponentials of the MS variables and their momenta (namely, the Weyl algebra), including linear operators, but on other physically relevant operators like e.g. the field Hamiltonian. It seems natural to demand that the MS Hamiltonian be well defined quantum mechanically (and be essentially self-adjoint). Additional conditions related with regularization may be important, although one would expect that regularization schemes should be a direct consequence of the quantization of the system, since this includes now the background, rather than having to import them from QFT, where one usually appeals to them claiming the absence of a quantization of the curved spacetime.

Assume then that we take a particular Fock representation for the MS modes in the family picked out by the commented criteria of vacuum invariance and unitary dynamics in the standard QFT regime. Let us call $\hat{a}_{\vec n,\epsilon}$ and $\hat{a}^\dagger_{\vec n,\epsilon}$ the corresponding annihilation and creation operators, acting on the Fock space ${\mathcal F}$, such that $[\hat{a}_{\vec n,\epsilon},\hat{a}^\dagger_{\vec n',\epsilon'}]=\delta_{\vec n,\epsilon}^{\vec n',\epsilon'}$. A basis of states is provided by the occupancy-number states
$|{\mathcal N} \rangle$, where ${\mathcal N}$ denotes an array of (positive integer) occupancy-numbers, one for each mode. The creation operator acting on these states excites the mode labeled by $\vec n$ and $\epsilon$, increasing the corresponding occupancy-number in one unit, as usual. 

\subsection{Quantum representation of the constraints}

In our gauge-invariant formulation the classical constraints Poisson commute, and therefore they can be imposed quantum mechanically without troubles as far as their quantum counterparts commute as well. We assume that our quantization satisfies this property and impose them independently.

Let us start with the linear perturbative constraints, which classically are $V^{\vec n,\epsilon}_{p_{\tilde l}}=0$ for $\tilde{l}=2,3$. These constraints are straightforward to impose adopting a quantization, for the part of the inhomogeneous sector parameterized by $\{V^{\vec n,\epsilon}_{{\tilde l}}\}$ (again with $\tilde{l}=2,3$), such that the operators representing the momenta act as derivatives (or as translations, in integrated or discrete versions) with respect to the configuration variables $V^{\vec n,\epsilon}_{q_{\tilde l}}$. Then, the constraints amount to require that physical states are independent of those variables. Hence, after imposing these constraints, we can just restrict the discussion to a representation space of the form $\mathcal{H}=\mathcal{H}_\text{kin}^\text{grav}\otimes \mathcal{H}_\text{kin}^\text{matt}\otimes {\mathcal F}$ to study all physically relevant quantum states. Let us emphasize that the restriction to these states is quantum mechanical, and not a classical reduction. Besides, let us note that these states are not yet fully physical, since we must still impose the zero-mode of the Hamiltonian constraint.

More challenging is the imposition of this global Hamiltonian constraint, for which we will only be able to provide approximate solutions. Let us first focus on its quantum representation. Classically this constraint is given by  
\begin{align}\label{dens}
 H_{|0}+ \breve H_{|2} \equiv e^{-3 {\tilde \alpha} }\tilde H=0,
\end{align}
with $H_{|0}$ defined as in Eq. \eqref{eq:H_0} but in terms of the background variables $(\tilde\alpha,\pi_{\tilde\alpha},\tilde\varphi,\pi_{\tilde\varphi})$, and $\breve H_{|2}=\sum_{\vec n,\epsilon} \breve H^{\vec n,\epsilon}_{|2}$ defined in Eq. \eqref{eq:H2MS}. In what follows we will focus on the densitized Hamiltonian constraint\footnote{We could also choose to work with the non-densitized constraint and carry out the densitization at the quantum level (see e.g. Ref. \cite{hybridFRWflatMS}).} $\tilde H=0$. 

Before continuing, let us introduce some convenient notation. We first define 
\begin{equation}
\mathcal H_0^{(2)}\equiv \pi_{\tilde\alpha}^2-e^{6\tilde\alpha}\tilde m^2 \tilde\varphi^2,\label{densH0}
\end{equation}
so that the contribution of the homogeneous sector to $\tilde H$ is $ e^{3 {\tilde \alpha} }H_{|0}= \big(\pi_{\varphi}^2-\mathcal H_0^{(2)}\big)/2$. In addition, we introduce the following functions of the zero-modes:
\begin{subequations}\label{eq:geometry-operators}
\begin{align}
\vartheta_o&\equiv  - 12 e^{4\tilde\alpha} \tilde m^2 \frac{\tilde\varphi}{\pi_{\tilde\alpha}}, \label{eq:geometry-operators1} \\
\!\vartheta_e &\equiv e^{2\tilde\alpha},\\
\!\vartheta_e^{q} &\equiv e^{-2\tilde\alpha} \pi_{\tilde\alpha}^2+\tilde{m}^2 e^{4{\tilde\alpha}}\left(1+15{\tilde\varphi}^2-18 e^{6{\tilde\alpha}}\tilde{m}^2 \frac{{\tilde\varphi}^4}{\pi_{\tilde\alpha}^2} \right) \nonumber\\
& = e^{-2\tilde\alpha} \mathcal H_0^{(2)} \left(19- 18 \frac{\mathcal H_0^{(2)}}{\pi_{\tilde\alpha}^2} \right)+\tilde{m}^2e^{4{\tilde\alpha}}\left(1-2{\tilde\varphi}^2 \right), \label{eq:geometry-operators3}
\end{align}
\end{subequations}
and we call
\begin{subequations}\label{eq:perturbation-operators}
\begin{align}
\Theta_o^{\vec n,\epsilon} &\equiv  - \vartheta_o (V^{\vec n,\epsilon}_{q_1})^2\\
\Theta_e^{\vec n,\epsilon} &\equiv -\left[(\vartheta_e \omega_n^2+\vartheta_e^q)(V^{\vec n,\epsilon}_{q_1})^2+ \vartheta_e (V^{\vec n,\epsilon}_{p_1})^2\right],\\
\Theta_o&\equiv \sum_{\vec n, \epsilon}\Theta_o^{\vec n,\epsilon},\qquad 
\Theta_e\equiv \sum_{\vec n, \epsilon}\Theta_e^{\vec n,\epsilon}.
\end{align}
\end{subequations}
Then, $2 e^{3\tilde{\alpha}}\breve H_{|2} = -( \Theta_e+ \Theta_o \pi_{\varphi}).$
Let us assume that we can represent the quantities introduced in Eqs. \eqref{densH0} and \eqref{eq:geometry-operators} as densely defined operators, $\hat{{\mathcal H}}_0^{(2)}$, $ \hat{\vartheta}_{o}$, $\hat{\vartheta}_{e}$, and $\hat{\vartheta}_{e}^q$, on $ \mathcal H_\mathrm{kin}^\mathrm{grav}\otimes\mathcal H_\mathrm{kin}^\mathrm{matt}$ (while acting as the identity on $\mathcal F$), and the objects given in Eqs. \eqref{eq:perturbation-operators} as densely defined operators, $\hat  \Theta_o$ and  $\hat  \Theta_e$, on $ \mathcal H_\mathrm{kin}^\mathrm{grav}\otimes\mathcal H_\mathrm{kin}^\mathrm{matt}\otimes \mathcal F$.
Then, we obtain the following operator representing $\tilde H$: 
\begin{align}\label{dens-constraint}
\hat {\tilde H}=\frac12 \left[\hat\pi_{\tilde\varphi}^2-\hat{\mathcal H}_0^{(2)}-\hat{\Theta}_e-\left(\hat\Theta_o\hat\pi_{\tilde\varphi}\right)_\text{S} \right],
\end{align}
where we have adopted the symmetrization
\begin{align}
\left(\hat\Theta_o\hat\pi_{\tilde\varphi}\right)_\text{S} =\frac12\left(\hat\Theta_o\hat\pi_{\tilde\varphi}+\hat\pi_{\tilde\varphi}\hat\Theta_o\right)=\frac12 [\hat\pi_{\tilde\varphi},\hat\Theta_o]+\hat\Theta_o\hat\pi_{\tilde\varphi}.
\end{align}

\subsection{Born-Oppenheimer ansatz}\label{BOa}
We will now analyze solutions to the (zero-mode of the) Hamiltonian constraint of the system by adopting the ansatz
\begin{equation}\label{BOans}
\Psi=\Gamma(\tilde\alpha,\tilde\varphi) \psi({\mathcal N},\tilde\varphi),
\end{equation}
where the dependence on the MS variables is denoted by the label ${\mathcal N}$ of the  occupancy-number states for the inhomogeneous subsector $\mathcal F$, and $\tilde\alpha$ denotes the dependence on the geometry of the homogeneous FRW sector.\footnote{This notation does not imply that we adopt a representation for the homogeneous geometry in which we have an operator representing $\tilde\alpha$ that acts by multiplication, but it is rather a symbolic way to indicate functional dependence on the homogeneous geometry sector.} Thus, the ansatz for the quantum states is that their wave functions $\Psi$ can be separated into two factors, one which depends on the  homogeneous degrees of freedom of the FRW geometry, and the other on the MS gauge-invariant modes. The ansatz allows for states that present different rates of variation in these two factors with respect to the zero-mode $\tilde\varphi$ of the matter scalar field, and when this happens the Hamiltonian constraint can be approximated and simplified. Hence the name of Born-Oppenheimer for the introduced ansatz. From this perspective, it is convenient to interpret $\tilde\varphi$ as an internal time for the system (at least in some intervals of the evolution). We recall that we are assuming a representation of the homogeneous matter sector such that the functions of $\tilde{\varphi}$ act as multiplicative operators on $\Psi$.

Let us for a moment consider the unperturbed FRW model. Its Hamiltonian constraint operator is proportional to $\hat\pi_{\tilde\varphi}^2-\hat{\mathcal H}_0^{(2)}$. In the evolution picture that we are employing, where $\tilde\varphi$ plays the role of the time, positive-frequency solutions of this constraint are given by 
\begin{align}\label{gamma}
 \Gamma(\tilde\alpha,\tilde\varphi)= \hat U(\tilde\alpha;\tilde\varphi)\chi(\tilde\alpha),
\end{align}
$\hat U$ being the corresponding unitary evolution operator. If $\hat{\mathcal H}_0^{(2)}$ is self-adjoint, then one can project on its positive part (P.P.) and take $\sqrt{\text{P.P.}(\hat{\mathcal H}_0^{(2)})}$ as the $\tilde\varphi$-dependent self-adjoint operator that generates the time evolution in Eq. \eqref{gamma}. To maintain the discussion more general, and since the self-adjointness of $\hat{\mathcal H}_0^{(2)}$ might not be guaranteed, or a straightforward definition of a square root for it might not be available, we simply assume that the FRW part of the quantum states has the form \eqref{gamma}, where the family of evolution operators $\hat{U}$ is such that there exists a self-adjoint operator $\hat{\tilde{\mathcal H}}_0$ satisfying $[\hat\pi_{\tilde\varphi},\hat U]  {\hat U^{-1} }=\hat{\tilde{\mathcal H}}_0$. Note that the concepts of self-adjointness and unitarity that we use here are those corresponding to the FRW-geometry part of our Hilbert space,\footnote{In a conventional Schr\"odinger representation in which $\hat\pi_{\tilde\varphi}=-i \partial_{\tilde\varphi}$ we have the standard unitary evolution operator  
\begin{equation}\label{chievolu}
\hat U(\tilde\alpha;\tilde\varphi)={\mathcal P}\left[\exp{\left(i\int^{\tilde\varphi}_{\tilde\varphi_0} d\tilde{\varphi}\,\hat{\tilde{\mathcal H}}_0(\tilde\alpha, \tilde \varphi)\right)}\right]\nonumber.
\end{equation}
The symbol ${\mathcal P}$ denotes {\it time} ordering with respect to $\tilde\varphi$. For simplicity, we have fixed the reduced Planck constant $\hbar$ equal to the unit.} $\mathcal{H}_\text{kin}^\text{grav}$. 
Actually, within our perturbative framework, it is natural to consider $\Gamma$ as an approximate solution to the Hamiltonian constraint of the homogeneous sector, and not necessarily an exact one. In fact, as we will see, $\Gamma$ will supply an approximate solution of the FRW model inasmuch as the square of $\hat{\tilde{\mathcal H}}_0$ is {\it{close}} to $\hat{\mathcal H}_0^{(2)}$ in the sense that $ (\hat{\tilde{\mathcal H}}_0)^2  -\hat{\mathcal H}_0^{(2)} + [\hat{\pi}_{\tilde\varphi}, \hat{\tilde{\mathcal H}}_0]$ is negligibly small on $\Gamma$.  

On the other hand, in Eq. \eqref{gamma} we normalize the state $\chi(\tilde\alpha)$ to the unit in ${\mathcal H}_\mathrm{kin}^\mathrm{grav}$. This part of the quantum state can be understood as the state of the FRW geometry at a fixed initial time $\tilde{\varphi}_0$ (and hence it is independent of $\tilde{\varphi}$). A natural election would be to choose a very semiclassical state $\chi$, so that it is highly peaked on a certain homogeneous geometry (and, if this is possible, that it remains peaked under the evolution dictated by $\hat{U}$).

Let us plug the ansatz \eqref{BOans} in the constraint equation\footnote{Since one expects that solutions do not belong to the kinematical Hilbert space, one should rather impose the constraint on some kind of generalized states $(\psi|$ with the adjoint action, in the form $(\psi| \hat {\tilde H}^{\dagger}=0$. We do not do so because we do not want to complicate the notation even more.} $\hat {\tilde H} \Psi=0$. Taking into account that
\begin{subequations}
\begin{align}
\hat{\pi}_{\tilde\varphi}\Psi&=\Gamma (\hat{\pi}_{\tilde\varphi}\psi)+(\hat{\tilde{\mathcal H}}_0\Gamma)\psi,\\
\hat{\pi}_{\tilde\varphi}^2\Psi&=\Gamma (\hat{\pi}_{\tilde\varphi}^2\psi)+2 (\hat{\tilde{\mathcal H}}_0\Gamma) (\hat{\pi}_{\tilde\varphi}\psi)+([\hat{\pi}_{\tilde\varphi},\hat{\tilde{\mathcal H}}_0]\Gamma)\psi+  \big\{(\hat{\tilde{\mathcal H}}_0)^2\Gamma\big\}\psi,
\end{align} 
\end{subequations}
the constraint can be rewritten as
\begin{align}\label{const}
&\Big\{\Big( (\hat{\tilde{\mathcal H}}_0)^2- \hat{\mathcal H}_0^{(2)} + [\hat{\pi}_{\tilde\varphi},\hat{\tilde{\mathcal H}}_0]\Big)\Gamma\Big\}\psi+2 (\hat{\tilde{\mathcal H}}_0\Gamma) (\hat{\pi}_{\tilde\varphi}\psi)+\Gamma (\hat{\pi}_{\tilde\varphi}^2\psi)-\frac12[\hat{\pi}_{\tilde\varphi}-\hat{\tilde{\mathcal H}}_0,\hat{\Theta}_o](\Gamma\psi)\nonumber\\
&-\big\{\hat{\Theta}_e+(\hat{\Theta}_o\hat{\tilde{\mathcal H}}_0)_\text{S}\big\}(\Gamma\psi)-\hat{\Theta}_o\big\{\Gamma(\hat{\pi}_{\tilde\varphi}\psi)  \big\}=0.
\end{align}
We note that,with our assumptions on the representation of the zero-mode of the scalar field, the operators $[\hat{\pi}_{\tilde\varphi},\hat{\tilde{\mathcal H}}_0]$ and $[\hat{\pi}_{\tilde\varphi},\hat{\Theta}_o]$  depend on $\tilde\varphi$, that acts by multiplication as an operator, but are independent of $\hat{\pi}_{\tilde\varphi}$. Besides, notice that the first term of the above equation is the correction coming from the consideration of wave functions $\Gamma$ that are not exact solutions to the constraint of the homogeneous sector (to see this, it suffices to make $\psi$ constant and the  operators $\hat{\Theta}_o$ and $\hat{\Theta}_e$ of the MS modes equal to zero).

Now, let us consider the approximation that, regarding the homogeneous FRW geometry, only the terms corresponding to the expectation values on $\Gamma$ are relevant. In other words, when taking the inner product of the left hand side of Eq. \eqref{const} with $\Gamma$ with respect to the FRW geometry (namely in $\mathcal{H}_\text{kin}^\text{grav}$), we disregard possible quantum transitions from $\Gamma$ to other states mediated by the action of the constraint. From Eq. \eqref{const} it is not difficult to see that, for this approximation to hold, we only need the following  operators to have small relative dispersions on the state $\Gamma$ for all values of $\tilde\varphi$ (or, in other words, the following  operators must be highly peaked on $\Gamma$ along the evolution in $\tilde\varphi$): i)
$\hat{\tilde{\mathcal H}}_0$, ii) $\hat{\vartheta}_{e}$, iii) $-\frac{i}{2}{\mathrm d}_{\tilde\varphi}\hat{\vartheta}_{o}+(\hat{\vartheta}_{o}\hat{\tilde{\mathcal H}}_0)_S+\hat{\vartheta}_{e}^q$, and iv)  $-i {\mathrm d}_{\tilde\varphi}\hat{\tilde{\mathcal H}}_0 +(\hat{\tilde{\mathcal H}}_0)^2-\hat{\mathcal H}_0^{(2)}$, where we have defined\footnote{In the conventional case $\hat{\pi}_{\tilde\varphi}=-i \partial_{\tilde\varphi}$, ${\mathrm d}_{\tilde\varphi}$ is the total derivative in the Heisenberg picture corresponding to a time evolution in $\tilde\varphi$ generated by $\hat{\tilde{\mathcal H}}_0$.} 
\begin{equation}
-i{\mathrm d}_{\tilde\varphi}\hat O\equiv [\hat{\pi}_{\tilde\varphi}-\hat{\tilde{\mathcal H}}_0,\hat O],
\end{equation}
for any operator  $\hat O$. In principle, the condition on the operator iv) may be satisfied with a suitable choice of $\hat{\tilde{\mathcal H}}_0$, in agreement with our comments above.

Assuming that the considered approximation is valid, we then get
\begin{align}\label{constrainBO}
\hat{\pi}_{\tilde\varphi}^2 \psi &+ \left(2 \langle \hat{\tilde{\mathcal H}}_0 \rangle_{\Gamma} - \langle \hat{\Theta}_{o} \rangle_{\Gamma}\right) \hat{\pi}_{\tilde\varphi}\psi\nonumber\\
&=\bigg[\langle \hat{\Theta}_{e}+ \big(\hat{\Theta}_{o} \hat{\tilde{\mathcal H}}_0\big)_S\rangle_{\Gamma}
+i \langle   {\mathrm d}_{\tilde\varphi}\hat{\tilde{\mathcal H}}_0 - \frac{1}{2}{\mathrm d}_{\tilde\varphi}\hat{\Theta}_{o} \rangle_{\Gamma} +\langle \hat{\mathcal H}_0^{(2)} - (\hat{\tilde{\mathcal H}}_0)^2 \rangle_\Gamma\bigg] \psi.
\end{align}
Here we have introduced the symbol $\langle   \hat O \rangle_\Gamma$ to denote the expectation value of the operator $\hat O$ on $\Gamma$ in the Hilbert space $\mathcal H_\mathrm{kin}^\mathrm{grav}$, namely, only with respect to the inner product of the FRW geometry. Note that, in general, the result is an operator acting on $\mathcal H_\mathrm{kin}^\mathrm{matt}\otimes \mathcal{F}$.

The above equation leads to a generalized Schr\"odinger equation for the evolution (in $\tilde\varphi$) of the inhomogeneities provided that the following conditions are satisfied:
\begin{itemize}
\item[a)] $\langle \hat{\Theta}_{o} \rangle_{\Gamma}$ has to be negligible as compared to $\langle \hat{\tilde{\mathcal H}}_0 \rangle_{\Gamma}$. This is valid as long as the quadratic contribution of the MS modes given by $\hat{\Theta}_{o}$ remains small when compared to the introduced generator of the $\tilde\varphi$-evolution in the FRW case, something that certainly fits in the perturbative scheme that we have adopted for the treatment of the inhomogeneities.
\item[b)] $\hat{\pi}_{\tilde\varphi}^2 \psi$ has to be negligible as well. The self-consistency of this assumption is checked in Appendix \ref{appb}.
\item[c)]  $i \langle{\mathrm d}_{\tilde\varphi}\hat{\tilde{\mathcal H}}_0 - \frac{1}{2}{\mathrm d}_{\tilde\varphi}\hat{\Theta}_{o}\rangle_\Gamma$ has to be negligible in comparison with $\langle \hat{\Theta}_{e}+ \big(\hat{\Theta}_{o} \hat{\tilde{\mathcal H}}_0\big)_S\rangle_\Gamma$. Otherwise, the Schr\"odinger equation would contain a term destroying the unitary evolution of the MS modes and which is not present in the classical field equations of these gauge invariants.
\end{itemize}
Indeed, if these three conditions are satisfied, we get the generalized Schr\"odinger equation
\begin{align}\label{schroMS1}
\hat{\pi}_{\tilde\varphi}\psi=\frac{\langle \hat{\Theta}_{e}+ \big(\hat{\Theta}_{o} \hat{\tilde{\mathcal H}}_0\big)_S\rangle_{\Gamma}+\langle \hat{\mathcal H}_0^{(2)} - (\hat{\tilde{\mathcal H}}_0)^2 \rangle_\Gamma}{2 \langle \hat{\tilde{\mathcal H}}_0 \rangle_{\Gamma}}\psi.
\end{align}
On top of the previous conditions, let us consider as well the requirement
\begin{itemize}
\item[d)] $ \langle \hat{\mathcal H}_0^{(2)} - (\hat{\tilde{\mathcal H}}_0)^2 \rangle_\Gamma$ has to be negligible in comparison with $\langle \hat{\Theta}_{e}+ \big(\hat{\Theta}_{o} 
\hat{\tilde{\mathcal H}}_0\big)_S\rangle_\Gamma$. This is natural to assume (once $ \langle{\mathrm d}_{\tilde\varphi}\hat{\tilde{\mathcal H}}_0\rangle_{\Gamma}$ has been ignored) if $\Gamma$ is an approximate solution of the unperturbed FRW model, as argued before.
\end{itemize}
If the four conditions a)--d) hold, then the Schr\"odinger equation simplifies to
\begin{align}\label{schroMS}
\hat{\pi}_{\tilde\varphi}\psi=\frac{\langle \hat{\Theta}_{e}+ \big(\hat{\Theta}_{o} \hat{\tilde{\mathcal H}}_0\big)_S\rangle_{\Gamma}}{2 \langle \hat{\tilde{\mathcal H}}_0 \rangle_{\Gamma}}\psi.
\end{align}
Correspondingly, the Hamiltonian generating the dynamics of the perturbations in the internal time $\tilde\varphi$ is given by ${\langle \hat{\Theta}_{e}+ \big(\hat{\Theta}_{o} \hat{\tilde{\mathcal H}}_0\big)_S\rangle_{\Gamma}}/{2 \langle\hat{\tilde{\mathcal H}}_0 \rangle_{\Gamma}}$. This Hamiltonian is just the MS Hamiltonian, with its dependence on the variables of the homogeneous geometry evaluated at the expectation values corresponding to the quantum state $\Gamma$, and divided by the expectation value of $\hat{\tilde{\mathcal H}}_0$ on it. 

We see that assuming the Born-Oppenheimer ansatz and introducing a controlled number of approximations allows us to recover standard QFT for the gauge-invariant perturbations, which propagate on an effective homogeneous geometry than can be regarded as {\it{dressed}}. The term {\it{dressed}} refers here to the fact that this geometry is not a classical one, but it retains the main quantum corrections to the geometry that the MS variables feel \cite{dressed}.

\subsection{Effective equations for the Mukhanov-Sasaki variables}
\label{MSequations}

Employing the Born-Oppenheimer ansatz and the approximation that the state $\Gamma$ remains highly peaked on the operators that encode the effect of the FRW geometry in the zero-mode of the Hamiltonian constraint (and hence, strictly speaking, without the need to introduce the rest of approximations that lead to a Schr\"odinger equation for the wave function of the MS modes), one can further derive effective classical equations for the MS variables. These effective dynamics treat the perturbations as classical, namely they replace the annihilation and creation operators of the gauge-invariant inhomogeneities by their classical counterpart. The assumption that this replacement is acceptable in order to get effective equations does not seem too stringent, because the zero-mode of the Hamiltonian constraint is quadratic in the MS gauge invariants. Since the functions of $\tilde\varphi$ act by multiplication, we can see the quantum dynamics of the MS modes ruled by that Hamiltonian constraint as corresponding to time-dependent harmonic oscillators, with  $\tilde\varphi$ playing the role of internal time. 

In the previous subsection we saw that the evolution of the MS variables is governed by Eq. \eqref{constrainBO}, which can be interpreted as the result of imposing a constraint of the form:
\begin{align}
\hat{\mathcal C}_{\mathrm{per}}=\hat \pi_{\tilde\varphi}^2+ D_{\Gamma}(\tilde\varphi) \hat \pi_{\tilde\varphi} + E_{\Gamma}(\tilde\varphi) - \Big\langle \hat{\Theta}_{e}+ \big(\hat{\Theta}_{o} \hat{\mathcal H}_{0}\big)_S- \frac{i}{2}{\mathrm d}_{\tilde\varphi}\hat{\Theta}_{o} \Big\rangle_{\Gamma} .
\end{align}
Here, $D_{\Gamma}$ and $E_{\Gamma}$ are two functions of $\tilde\varphi$ which depend on the state $\Gamma$ of the homogeneous geometry, and which we do not specify because they are irrelevant for our calculations. As for the momentum $\hat \pi_{\tilde\varphi}$, it is supposed to act as a generalized derivative with respect to $\tilde\varphi$, and we neither need to specify it at this level of discussion. The constraint operator $\hat{\mathcal C}_{\mathrm{per}}$ is imposed on the sector of the model composed by $\mathcal H_\mathrm{kin}^\mathrm{matt}\otimes\mathcal F$.
The corresponding classical evolution generator, denoted by ${\mathcal C}_{\mathrm{per}}$, is then obtained by replacing $\hat \pi_{\tilde\varphi}$ by its explicit form as a generalized derivative, and the MS operators $\hat V^{\vec n,\epsilon}_{q_1}$ and $\hat V^{\vec n,\epsilon}_{p_1}$ by their classical counterparts, according to our comments above.

Taking into account the densitization of the constraint [set in Eq. \eqref{dens}] and the definition of the homogeneous part of the lapse function, one can check that ${\mathcal C}_{\mathrm{per}}/2$ generates reparameterizations in a time $\bar T$ that, at leading perturbative order, is related with the proper time by $dt=\sigma e^{3\alpha} d\bar T$.
Moreover, we can change to a conformal time  $\eta_{\Gamma}$, adapted to the {\it{dressed}} FRW geometry associated with the state $\Gamma$. Observing expressions \eqref{eq:perturbation-operators}, we see that all the dependence of ${\mathcal C}_{\mathrm{per}}/2$ on the MS momenta $V^{\vec n,\epsilon}_{p_1}$ is given by the term $\langle\hat \vartheta_e\rangle_{\Gamma}(V^{\vec n,\epsilon}_{p_1})^2/2$ coming from $\langle \hat{\Theta}_{e}\rangle_{\Gamma}$. Then, it is natural to define\footnote{It is possible to see that, with our choice of numeric factors, this definition of conformal time is not sensitive to the choice of $l_0$.}  
$l_0d{\eta_{\Gamma}}=\langle\hat \vartheta_e\rangle_{\Gamma} d\bar T$.
This change of time is well defined because $\langle\hat \vartheta_e\rangle_{\Gamma}$ is a c-number (depending on $\tilde\varphi$ through  $\Gamma$), and it is monotonous inasmuch as the operator $\hat \vartheta_e$ representing $\vartheta_e=e^{2\tilde\alpha}$ should be positive. It is worth emphasizing that we would have failed to define a change of time parameter had the time derivative $d{\eta_{\Gamma}}/d\bar T$ been an operator. Hence, the expectation value on $\Gamma$ is essential in order to introduce the above change of time. We also point out that the change is state dependent, and hence the properties of the evolution in the times $\bar T$ and $\eta_{\Gamma}$ can be quite different when considered in the physical Hilbert space of the system.

In summary, the evolution of $V^{\vec n,\epsilon}_{1}$ in the time $\eta_{\Gamma}$ is given (under Poisson brackets) by 
\begin{align}
d_{\eta_{\Gamma}} V^{\vec n,\epsilon}_{1}=\frac{l_0}{2\langle\hat \vartheta_e\rangle_{\Gamma}}\{V^{\vec n,\epsilon}_{1},{\mathcal C}_{\mathrm{per}}\},
\end{align}
where $d_{\eta_{\Gamma}}$ denotes the derivative with respect to $\eta_{\Gamma}$. The evolution of $V^{\vec n,\epsilon}_{q_1}$ is simply given by $d_{\eta_{\Gamma}} V^{\vec n,\epsilon}_{q_1}=l_0 V^{\vec n,\epsilon}_{p_1}$. Using this result and taking again the time derivative we get the following effective MS equations:
\begin{align}\label{MSeqhybrid}
d^2_{\eta_{\Gamma}}V^{\vec n,\epsilon}_{q_1}=- V^{\vec n,\epsilon}_{q_1} \left[\tilde \omega_n^2 +
\langle \hat{\theta}_{e}+ \hat{\theta}_{o}\rangle_{\Gamma}\right],
\end{align}
where we have defined $\tilde \omega_n^2=l_0^2 \omega_n^2$ and
\begin{subequations}
\begin{align}
\langle \hat{\theta}_{e}\rangle_{\Gamma}\equiv l_0^2 \frac{  \langle \hat{\vartheta}_{e}^q\rangle_{\Gamma}}{ \langle \hat{\vartheta}_{e}\rangle_{\Gamma}},\qquad \langle \hat{\theta}_{o}\rangle_{\Gamma}\equiv l_0^2 \frac{  \langle (\hat{\vartheta}_{o}\hat{\tilde{\mathcal H}}_0)_S-\frac{i}{2}\mathrm d_{\tilde\varphi}\hat{\vartheta}_{o}\rangle_{\Gamma}}{ \langle \hat{\vartheta}_{e}\rangle_{\Gamma}}.
 \end{align}
\end{subequations}
We note that the last term in the square brackets of this MS equation is a function of only $\tilde\varphi$, and hence of time, when the scalar field is evaluated on the solutions to the effective equations. This factor contains quantum modifications with respect to the standard MS equation, which encode the most relevant quantum geometry effects of the homogeneous background. The derived equations are still of harmonic oscillator type with time-dependent frequencies. Besides, no dissipation term appears and the equations are hyperbolic in the ultraviolet regime, where $\tilde \omega_n^2$ dominates in the square brackets.
We have included the contribution of $\big\langle \hat{\theta}_{o} \big\rangle_{\Gamma}$, although, in view of our discussion in Appendix \ref{appb}, we expect it to be negligible in practice as it is proportional to $\tilde m^2$, in the studied case of a mass term as the potential.

Let us say that, although in our analysis we have assumed for simplicity that the potential of the scalar field is given by a mass contribution, actually we can easily extend the discussion to a general potential $W(\tilde\varphi)$. For that, we simply need to replace in Eq. \eqref{eq:geometry-operators3} the two powers of $\tilde m^2\tilde\varphi^2$ with $2W(\tilde\varphi)$, and the term $\tilde m^2$ with $W''(\tilde\varphi)$, whereas the factor $\tilde m^2\tilde\varphi$ in Eq. \eqref{eq:geometry-operators1} has to be replaced with $W'(\tilde\varphi)$ (here the prime denotes derivative with respect to $\tilde\varphi$). Then, for the approximations related with the Born-Oppenheimer ansatz to hold, instead of requiring a small mass, we need to require small variations of the potential.

\section{LQC representation for the homogeneous sector}
\label{sec:lqc}

So far in our discussion we have left unspecified the concrete quantization adopted in the FRW-geometry part of the homogeneous sector. For the sake of illustrating our analysis with an example of particular interest, in this section we will adopt the polymeric quantization employed in LQC. There are plenty of references where the details of the polymeric quantization of the unperturbed FRW model can be found (see e.g. \cite{lqc,bounce1,bounce2,mmo}). More specifically, we will adhere to the so-called {\it improved dynamics} prescription of LQC \cite{bounce2} and, among the possible symmetric orderings for the Hamiltonian constraint operator, we will adopt the prescription put forward in Ref. \cite{mmo}, as it proves to be  most convenient both for theoretical and practical purposes \cite{pres}.
Moreover, in addition, for the homogeneous scalar field, we will use the standard Schr\"odinger quantization which is usually taken in LQC. Besides, in the perturbed model, when representing the operators of the homogeneous sector that are coupled to perturbations, we will follow Ref. \cite{hybridFRWflatMS}, that adopts the factor ordering of Ref. \cite{mmo} as well.

In the loop formalism, the geometry is described by the Ashtekar-Barbero $su(2)$ connection and by the densitized triad \cite{lqg}, that form a canonical pair. In FRW cosmologies, owing to homogeneity and isotropy, they are respectively determined simply by two dynamical variables, $c$ and $p$, with Poisson bracket equal to $8\pi G\gamma/3$, where $\gamma$ is the Immirzi parameter \cite{Immirzi}. The mentioned improved dynamics scheme accounts for the existence of a minimum non-vanishing eigenvalue $\Delta$ of the area operator in LQG. This scheme involves a transformation to the new variables
\begin{align}\label{vp}
v=\sgn{(p)} \frac{|p|^{3/2}}{2\pi G \gamma  \sqrt{\Delta}},\quad b=\sqrt{\frac{\Delta}{|p|}}c\equiv \bar\mu c.
\end{align}
The sign of $p$ determines the orientation of the triad. The new variables verify $\{b, v\}=2$. The variable $v$ is proportional to the volume of the homogeneous model, which is finite for the three-torus spatial topology under study, and given by $V=2\pi G \gamma  \sqrt{\Delta}|v|$.
These variables for the homogeneous geometry are related with those of previous sections, $(\tilde\alpha,\pi_{\tilde\alpha})$, via the canonical transformation
\begin{align}
e^{\tilde\alpha}=\left( \frac{3\gamma \sqrt{\Delta}}{2\sigma}|v|\right)^{1/3}, \qquad \pi_{\tilde\alpha}=-\frac32 v b.
\end{align}

On the other hand, for the zero-mode of the matter scalar field, it is convenient to introduce the following scaling by a constant, $\phi = \tilde \varphi/(l_0^{3/2}\sigma)$ and $\pi_\phi = l_0^{3/2}\sigma\pi_{\tilde \varphi}$, since this is the usual parameterization employed in the LQC literature.

In LQC one {\it{polymerizes}} the connection, something which means that the connection coefficient $c$ has no well-defined operator in the quantum theory but one represents instead its holonomy elements, given by the exponentials of $b$.  
As a consequence, the FRW-geometry sector of the kinematical Hilbert space, $\mathcal{H}_\text{kin}^\text{grav}$, is the span of basis states $|v\rangle$, with $v\in\mathbb{R}$, normalized with respect to the discrete inner product $\langle v'|v\rangle=\delta_{v}^{v'}$. The operator $\hat{v}$ , which acts by multiplication,
$\hat{v} |v\rangle=v  |v\rangle$, has a discrete spectrum. We are taking the reduced Planck constant $\hbar$ equal to one (as in Sec. \ref{BOa}) to simplify the notation.  The holonomy operators $N_{\pm\bar\mu}\equiv e^{\pm i\bar\mu c/2}= e^{\pm i b/2}$ produce a constant shift in the label of these states, $ \hat N_{\pm\bar\mu} |v\rangle=|v\pm1\rangle$, as one can deduce from the commutation relations  $[ \hat N_{\bar\mu},\widehat{v}]=i\widehat{\{N_{\bar\mu},v}\}$.
As a result of this loop representation, the classical expression $cp=2\pi G \gamma v b$  gets promoted in the quantum theory to a symmetric version of $\hat{v}\,\widehat{\sin(b)}$ multiplied by $2\pi G \gamma$, where $\widehat{\sin(b)}=i( \hat N_{-2\bar\mu}- \hat N_{2\bar\mu})/2$. We choose the symmetric ordering  proposed in Ref. \cite{mmo}, after which the operator representing $cp$ becomes
\begin{align}\label{eq:Omega}
\hat{\Omega}_0\equiv\frac1{2\sqrt{\Delta}}{\hat V}^{1/2}\left[\widehat{\text{sgn}(v)}\widehat{\sin(b)}+\widehat{\sin(b)}\widehat{\text{sgn}(v)}\right]{\hat V}^{1/2}.
\end{align} 

As we have said, for the zero-mode of the scalar field, we adopt a standard Schr\"odinger representation, with $\hat{\pi}_\phi=i\partial_\phi$ and $\hat\phi$ acting by multiplication, so that $\mathcal H_\mathrm{kin}^\mathrm{matt}=L^2(\mathbb{R},d\phi)$.
In total, this LQC representation yields the following expression for the operator $\hat{\mathcal H}_0^{(2)}$ [see Eqs. \eqref{densH0} and \eqref{dens-constraint}]:
\begin{align}\label{eq:calH_0}
\hat{\mathcal H}_0^{(2)} &=\frac{3}{4\pi G}\left( \frac{3}{4\pi G \gamma^2}\hat\Omega_0^2- \hat V^2 m^2\hat\phi^2\right).
\end{align}
The operator  $\hat\Omega_0^2$ annihilates the zero-volume state $| v=0 \rangle$ and leaves invariant its orthogonal complement. Moreover, it leaves invariant the subspaces $\mathcal H_\mathrm{\varepsilon}^\pm$ formed by states with support on the semilattices $\mathcal L_\mathrm{\varepsilon}^\pm=\{\pm(\varepsilon+4n)|n\in\mathbb N\}$, where $\varepsilon\in(0,4]$. $\mathcal H_\mathrm{\varepsilon}^\pm$ are separable, in contrast with the original $\mathcal H_{\mathrm{kin}}^{\mathrm{grav}}$. Notice also that, in each of these sectors, the homogeneous volume $v$ has a strictly positive minimum $\varepsilon$ (or negative maximum $-\varepsilon$). Below we will represent the quadratic contributions of the inhomogeneities to the (zero-mode of) the Hamiltonian constraint by operators that also leave these semilattices invariant, that therefore get superselected. In the following, we will restrict the discussion, e.g., to $\mathcal H_\mathrm{\varepsilon}^+$, spanned by states with positive $v
 \in \mathcal L_\mathrm{\varepsilon}^+$.

Let us consider now the representation of the homogeneous terms entering the quadratic contribution of the inhomogeneities to the zero-mode of the Hamiltonian constraint, namely the operators  $ \hat{\vartheta}_{o}$, $\hat{\vartheta}_{e}$, and $\hat{\vartheta}_{e}^q$ representing the $\vartheta$-terms in Eqs. \eqref{eq:geometry-operators}. 
These terms are affected in principle by some quantization ambiguities. As it was done in Refs. \cite{hybridFRWflat,hybridFRWflatMS}, we will introduce a symmetric factor ordering that tries to respect, as far as possible, the assignations of representation made in the homogeneous sector of the system as follows. i) We represent the inverse of the volume with the standard regularization of LQC, namely $\widehat{[1/V]}$  is the cube of the regularized operator
\begin{align}
\widehat{\left[\frac1{V}\right]}^{1/3}=\frac3{4\pi G \gamma \sqrt{\Delta}}\widehat{\text{sgn}(v)} \hat V^{1/3}\left[\hat N_{-\bar\mu}\hat V^{1/3}\hat N_{\bar\mu}-\hat N_{\bar\mu}V^{1/3}\hat N_{-\bar\mu}\right].
\end{align}
This operator annihilates the state $|v=0\rangle$ and commutes with $\hat V$, and hence it is well defined on the subspaces $\mathcal H_\mathrm{\varepsilon}^\pm$.
ii) The products of the form $f(\phi)\pi_{\phi}$, where $f$ is an arbitrary function, are represented with the symmetric factor ordering $\big(f(\hat \phi) \hat\pi_\phi+\hat\pi_\phi f(\hat\phi)\big)/2$. iii) We adopt an algebraic symmetrization in factors of the form $V^rg(cp)$, that are promoted to the operators $\hat V^{r/2}\hat g\hat V^{r/2}$, where $g$ is any function and $r$ a real number. This algebraic symmetric
factor ordering is also adopted for powers of the inverse volume. iv) Even powers of  $\Omega_0\equiv -l_0^3\sigma^2 \gamma \pi_{\tilde\alpha}$ are represented by the same powers of the operator $\hat \Omega_0$, as in FRW. And v) in the case of odd powers of $\Omega_0$, let us say $\Omega_0^{2k+1}$ with $k$ equal to an integer, we adopt the representation $|\hat\Omega_0|^k \hat\Lambda_0|\hat\Omega_0|^k$, where $|\hat\Omega_0|$ is the square root of the positive operator $\hat\Omega_0^2$ and $\hat{\Lambda}_0$ is defined exactly as $\hat \Omega_0$ in Eq. \eqref{eq:Omega}, but with holonomies of double length. In this way, its action only shifts $v$ in multiples of four units, and hence preserves the superselection sectors of the homogeneous geometry. Following these prescriptions, we obtain the operators
\begin{subequations}
 \begin{align}
\hat\vartheta_o&= \frac{4}{l_0} \sqrt{12\pi G} \gamma m^2\hat\phi \hat V^{2/3} |\hat\Omega_0|^{-1} \hat\Lambda_0|\hat\Omega_0|^{-1}\hat V^{2/3} ,\\
\hat\vartheta_e&=\frac{3 l_0}{4\pi G}\hat V^{2/3},\\
\hat\vartheta_e^q&=\frac{4\pi G}{3 l_0}\widehat{\left[\frac1{V}\right]}^{1/3}\hat{\mathcal H}_0^{(2)}\left(19-32 \pi^2 G^2 \gamma^2 \hat\Omega_0^{-2}\hat{\mathcal H}_0^{(2)}\right) \widehat{\left[\frac1{V}\right]}^{1/3} \nonumber\\
& +\frac{3 m^2}{4 \pi G l_0 }\hat V^{4/3}\left( 1- \frac{8\pi G}{3} \hat\phi^2\right),
\end{align}
\end{subequations}
which are densely-defined on $\mathcal H_\mathrm{\varepsilon}^+\otimes L^2(\mathbb{R},d\phi)$. These results coincide with those of Ref. \cite{hybridFRWflatMS} except for some factors of the inverse of the volume operator, which appear now as the inverse powers of the volume operator itself. This difference arises from our different choice of densitization for the zero-mode of the Hamiltonian constraint at the classical level (the densitization was done at the quantum level in that mentioned, previous work). 

As we have already said, the validity of the Born-Oppenheimer approximation depends both on the particular representation chosen for the above homogeneous operators, and on the properties of the homogeneous states $\Gamma$. We refer the reader to Ref. \cite{hybridFRWflatMS} for more comments at this respect regarding the loop quantization.

\section{Conclusions}
\label{sec:conclusions}

In this work we have developed a covariant formulation of the gravitational system that describes a perturbed flat FRW spacetime with a compact three-torus topology and minimally coupled to a scalar field. In our perturbative scheme, we truncate the action of the model at quadratic order in the perturbations, and focus our attention on scalar perturbations. We expand the spatial dependence in Fourier modes, using a basis of eigenfunctions of the Laplace-Beltrami operator of the spatial sections. Besides, we treat the degrees of freedom of the zero-modes exactly, up to the order of our truncation. The total Hamiltonian of the resulting model is a sum of constraints, formed by a linear combination of the zero-mode of the scalar (or Hamiltonian) constraint, and of all the inhomogeneous modes of two local constraints which are linear in the perturbations. One of them comes from the perturbations of the Hamiltonian constraint and the other corresponds to the perturbation of the momentum (or diffeomorphisms) constraint. The zero-mode of the Hamiltonian constraint is, in turn, formed by the constraint of the unperturbed model plus contributions that are quadratic in the perturbations.

The first important aspect of our treatment is that we do not fix the gauge freedom associated with the linear perturbative constraints. In order to deal with them, we abelianize their algebra. We achieve this by replacing the linear perturbative Hamiltonian constraint with a suitable linear combination of it and of the constraint of the unperturbed model. At the perturbative order of our truncation of the action, this replacement amounts to a redefinition of the zero-mode of the lapse function. As a result of this process, we are able to parameterize the perturbations of the model with these Abelian linear perturbative constraints and the modes of the MS field, together with their corresponding canonically conjugate variables. Since the MS modes Poisson commute with the constraints, we indeed attain a parameterization for the perturbations fully adapted to gauge invariance. 

Another goal of our work consists in extending the canonical transformation from the perturbations to the entire system, namely considering not only the inhomogeneities, but also the homogeneous sector of the model (the zero-modes describing the degrees of freedom present in the unperturbed FRW model). In this way we get at our level of truncation in the perturbations a fully covariant description of the entire symplectic manifold, and not just of its inhomogeneous sector formed by the perturbations, something that would have required treating the zero-modes as fixed, rather than as genuine degrees of freedom. This is an important step towards the quantization of the system, inasmuch as without this completion of the canonical transformation, the system would have lost its symplectic structure. As a result of this canonical transformation, the zero-mode of the lapse gets redefined again, absorbing a quadratic contribution of the perturbations, but keeping the original freedom as an unspecified Lagrange multiplier. In addition, also the Lagrange multipliers accompanying the perturbative linear constraints get redefined similarly (though in this case with linear contributions of the perturbations), so that the final Hamiltonian is a linear
combination of (first-class) commuting constraints. We have derived the explicit expression of the spacetime metric in terms of our gauge-invariant phase-space variables and new Lagrange multipliers. 

In our construction, we have started with a form of the linear perturbative constraints that has been suitably scaled by powers of the scale factor. It is worth explaining that we might have started as well without this scaling of the perturbative constraints: The difference would have simply led to another definition of the momentum of the zero-mode of the FRW geometry in the final canonical set of variables for the entire system. 

Once the classical description has been completed, we have proceeded to quantize the system. For this, we have proposed a hybrid approach that combines any quantization of the homogeneous FRW sector of the model with a more standard field description of the inhomogeneous degrees of freedom. The utility of this approach rests on the assumption that the most relevant quantum geometry effects are those affecting the zero-modes of the system, and therefore may require a more careful analysis, while perturbations, even being quantum, can be described in a more conventional way. In other words, the philosophy behind the hybrid quantization consists in adopting a representation for the homogeneous sector of the model capable to capture its quantum gravity nature, while the perturbations are treated essentially along the lines of QFT in curved spacetimes.

We have followed the Dirac approach of imposing the constraints as quantum operators in order to find physical states of the system. As the classical constraints Poisson commute, it is natural to pass to quantum counterparts that commute as well, and hence we can impose them independently. The linear perturbative constraints generate translations in their canonically conjugate variables. Their quantum imposition then means that physical states do not depend at all on those gauge degrees of freedom. Therefore, thanks to our abelianization of the constraint algebra, we see that the process of imposing the perturbative linear constraints in the quantum theory leads to quantum states that depend only on the MS modes and the homogeneous sector. Remarkably, this is precisely the kind of states obtained with the approach adopted e.g. in Ref. \cite{hybridFRWflatMS}, even if the avenue followed in that case included a gauge fixing of the perturbations, although the description of the remaining physical degrees of freedom was done only in terms of gauge invariants. In this sense our results, in the present work, support those obtained with gauge fixing and show the covariance of the conclusions attained previously, at the considered perturbative level. After the imposition of the linear perturbative constraints, the only constraint on the model is the zero-mode of the Hamiltonian constraint.

To solve this Hamiltonian constraint, we have adopted an ansatz for physical quantum states of Born-Oppenheimer type, and discussed the validity of such approximation. This Born-Oppenheimer ansatz has a similar motivation as the one proposed in Ref. \cite{vidotto}. It regards the zero-mode of the scalar field as an internal clock and assumes that the MS modes do not affect much the motion of the zero-modes of the geometry. The ansatz assumes a separation of the quantum dependence on the FRW geometry and on the MS gauge invariants. If the state of the FRW geometry remains highly peaked with respect to three operators [the operators listed as i)-iii) in the paragraph above Eq. \eqref{constrainBO}] when it changes in the internal time, and this evolution fits suitably that of the unperturbed system, the zero-mode of the Hamiltonian constraint provides a quantum dynamical equation for the MS variables in terms of this time. Furthermore, under certain conditions, the ansatz allows one to introduce a series of approximations that lead to a Schr\"odinger equation for the MS modes, characterized by a Hamiltonian that is given by the MS Hamiltonian corrected by the fact that its dependence on the homogeneous geometry gets {\it dressed} with quantum corrections. In this manner, one recovers a QFT for the gauge-invariant perturbations, that propagate in a fixed (and in general non-classical) spacetime. 

Generically, when one is dealing with quantum fields in curved spacetimes, there is ambiguity both in the way that one parameterizes the field (since one can introduce scalings by background functions) and in the Fock representation chosen for that field. Remarkably, for the system under study, there exist uniqueness theorems that pick out a preferred parameterization (in fact, a canonical pair for the field) and a class of unitarily equivalent Fock representations for it \cite{uniqueness1,uniqueness2,uniquenessother,uniqueperturb}. These choices are selected by the natural criteria of having a vacuum invariant under the symmetries of the background and a unitary implementation of the dynamics. Our parameterization for the MS modes matches precisely this description.\footnote{Nonetheless, we emphasize that our strategy of a hybrid approach to the quantization can be easily adapted to other parameterizations.} No regime with a unitary QFT would have been accessible if we had chosen a different parameterization for the MS field (by means of a time-depending scaling).

Another important conclusion of our analysis is the robustness of the class of dynamical equations that govern the evolution of the MS gauge invariants. We have shown that the dynamics of these invariants are always given by a harmonic oscillator equation, with a frequency that depends on the internal time, as long as the Born-Oppenheimer states are highly peaked with respect to the operators that feel the FRW geometry in the zero-mode of the Hamiltonian constraint, and irrespectively of whether this constraint equation can be approximated by one of Schr\"odinger type. To arrive at this conclusion, essentially our only additional assumption has been that the effective dynamics of the MS modes can be derived with the substitution of annihilation and creation operators by their counterparts as classical variables. The MS equation obtained in this manner contains quantum modifications to the time-dependent part of the frequency, but the modification is the same for all modes, and in particular it does not affect the ultraviolet behavior of the classical equations for these perturbations in general relativity. 

Let us also note that, in order to obtain the quantum evolution for the perturbations, we have not employed a semiclassical approximation for the homogeneous geometry. We do not even need to exactly solve the unperturbed model to determine the quantum state of the background geometry. Indeed, within our perturbative scheme, it is enough that we consider approximate solutions, whose difference with respect to the exact ones can be neglected perturbatively.

We finish by emphasizing that, if we believe in QFT in curved spacetimes, it is natural to think that there is a deeper quantum regime, before reaching a complete quantum gravity description, in which our hybrid quantization makes sense. This hybrid approach  encodes the main quantum gravity effects and it is suitable to potentially predict whether there exist modifications of quantum gravity nature regarding cosmological observables, for instance in the power spectrum of the CMB. It therefore offers a framework to extract physical consequences of quantum gravity in cosmology.

\acknowledgments
The authors are greatly thankful to J. Olmedo for enlightening conversations, assistance in computations with {\it Mathematica}, and pointing out some important references. In addition, they are grateful to B. Elizaga Navascu\'es, T. Pereira, and S. Tsujikawa for discussions. This work was partially supported by the Spanish MICINN/MINECO Project No. FIS2011- 30145-C03-02 and its continuation FIS2014-54800-C2-2-P. M. M-B acknowledges financial support from the Netherlands Organisation for Scientific Research (NWO) (Project No. 62001772).

\appendix
\section{Spacetime metric and scalar field in the gauge-invariant formulation}
\label{app}

\subsection{Zero-modes}

In this appendix, we give the explicit expressions of the original canonical variables for the homogeneous sector of the phase-space of our system, $\{w^a\}=\{\alpha, \pi_{\alpha}, \varphi, \pi_{\varphi}\}$, in terms of the new canonical set introduced in the text, formed by the variables $\tilde w^a$ and $ V^{\vec n,\epsilon}_{l}$.
Substituting Eqs. \eqref{eq:transf} and their corresponding derivatives into Eqs. \eqref{eq:transf-background} yields the desired expressions:
\begin{subequations}
\begin{align}\label{alphapp}
\alpha&=\tilde\alpha -\frac12 \sum_{\vec n, \epsilon} \Bigg[ e^{-2\tilde\alpha} \left(1-3\frac{\pi_{\tilde\varphi}^2}{\pi_{\tilde\alpha}^2}\right) (V^{\vec n,\epsilon}_{q_1})^2 + 2 \frac{e^{-\tilde\alpha}}{\pi_{\tilde\alpha}} \left(e^{6\tilde\alpha}\tilde{m}^2 \tilde\varphi + 3\frac{\pi_{\tilde\varphi}^3}{\pi_{\tilde\alpha}}+\pi_{\tilde\alpha}\pi_{\tilde\varphi}\right)V^{\vec n,\epsilon}_{q_1} V^{\vec n,\epsilon}_{q_2}\Bigg]\nonumber\\
&-\frac12 \sum_{\vec n, \epsilon} \Bigg[  \left(\pi_{\tilde\alpha}^2+3\pi_{\tilde\varphi}^2+\frac13 e^{4\tilde\alpha} \omega^2_{\vec n} \right) (V^{\vec n,\epsilon}_{q_2})^2+\frac23 \pi_{\tilde\alpha} V^{\vec n,\epsilon}_{q_2} V^{\vec n,\epsilon}_{q_3}+ 2  e^{\tilde\alpha} \frac{\pi_{\tilde\varphi}}{\pi_{\tilde\alpha}} V^{\vec n,\epsilon}_{q_2} V^{\vec n,\epsilon}_{p_1}\Bigg] \nonumber\\&+\frac12 \sum_{\vec n, \epsilon} \Bigg[\frac2{\pi_{\tilde\alpha}}V^{\vec n,\epsilon}_{q_2} V^{\vec n,\epsilon}_{p_2}+\frac13 (V^{\vec n,\epsilon}_{q_3})^2 \Bigg],\\
\pi_\alpha&=\pi_{\tilde\alpha}- \sum_{\vec n, \epsilon} \Bigg[ 6 e^{5\tilde\alpha}\tilde{m}^2 \tilde\varphi V^{\vec n,\epsilon}_{q_1} V^{\vec n,\epsilon}_{q_2}+\left( 3e^{6\tilde\alpha}\tilde{m}^2 \tilde\varphi \pi_{\tilde\varphi}+\frac23 \pi_{\tilde\alpha}e^{4\tilde\alpha} \omega^2_{\vec n}\right) (V^{\vec n,\epsilon}_{q_2})^2\Bigg]\nonumber\\
&+ \sum_{\vec n, \epsilon} V^{\vec n,\epsilon}_{q_1} V^{\vec n,\epsilon}_{p_1},
\end{align}
\end{subequations}
\begin{subequations}
	\begin{align}\label{phiapp}
\varphi&=\tilde\varphi + \sum_{\vec n, \epsilon} \Bigg[ e^{\tilde\alpha}V^{\vec n,\epsilon}_{p_1} V^{\vec n,\epsilon}_{q_2}  +e^{-\tilde\alpha}\left(\pi_{\tilde\alpha}+3\frac{\pi_{\tilde\varphi}^2}{\pi_{\tilde\alpha}}\right) V^{\vec n,\epsilon}_{q_1} V^{\vec n,\epsilon}_{q_2}+\left(3\pi_{\tilde\alpha}\pi_{\tilde\varphi}-\frac{1}{2} e^{6\tilde\alpha}\tilde{m}^2 \tilde\varphi\right)(V^{\vec n,\epsilon}_{q_2})^2\Bigg]\nonumber\\
&+ \sum_{\vec n, \epsilon} \Bigg[e^{-\tilde\alpha} V^{\vec n,\epsilon}_{q_1} V^{\vec n,\epsilon}_{q_3}+ \pi_{\tilde\varphi}  V^{\vec n,\epsilon}_{q_2} V^{\vec n,\epsilon}_{q_3}-3 e^{-2\tilde\alpha} \frac{\pi_{\tilde\varphi}}{\pi_{\tilde\alpha}}(V^{\vec n,\epsilon}_{q_1})^2\Bigg],\\
\pi_\varphi&=\pi_{\tilde\varphi} -\sum_{\vec n, \epsilon} \Bigg[e^{5\tilde\alpha}\tilde{m}^2 V^{\vec n,\epsilon}_{q_1} V^{\vec n,\epsilon}_{q_2}+\frac12 e^{6\tilde\alpha}\tilde{m}^2\pi_{\tilde{\varphi}}(V^{\vec n,\epsilon}_{q_2})^2\Bigg].
\end{align}
\end{subequations}

\subsection{Lapse function and shift vector}

The reformulation of the model in terms of gauge invariants for the perturbations leads to the Hamiltonian given in Eq. \eqref{eq:Hamiltonian3}, with Lagrange multipliers $\{\bar N_0,  G_{\vec n,\epsilon},  K_{\vec n,\epsilon}\}$, and constraints expressed in terms of the new zero-mode variables $\{\tilde\alpha, \pi_{\tilde\alpha}, \tilde\varphi, \pi_{\tilde\varphi}\}$ and new variables for the inhomogeneities $\{V^{\vec n,\epsilon}_{q_l},V^{\vec n,\epsilon}_{p_l}\}$. Comparing Eq. \eqref{eq:Hamiltonian3} with Eq. \eqref{eq:Hamiltonian2}, and taking Eq. \eqref{eq:newH2} into account, we get (within our perturbative truncation scheme)
\begin{subequations}
\begin{align}
G_{\vec n,\epsilon}&=g_{\vec n,\epsilon}+N_0  F_{|1}^{\vec n,\epsilon}, \\ 
\bar N_0 &=N_0+\sum_{\vec n,\epsilon}\big(3 e^{3\tilde{\alpha}}g_{\vec n,\epsilon}a_{\vec n,\epsilon}+N_0 F_{|2}^{\vec n,\epsilon}\big) ,\\
K_{\vec n,\epsilon}&=k_{\vec n,\epsilon}+N_0 \bigg(F_{\_1}^{\vec n,\epsilon}-3 \frac{e^{-3{\tilde\alpha}}}{\pi_{\tilde\alpha}} V^{\vec n,\epsilon}_{p_2} + \frac92e^{-3{\tilde\alpha}} V^{\vec n,\epsilon}_{p_3}\bigg).
\end{align}
\end{subequations}
From these expressions we deduce the inverse relations (again within the order of our perturbative truncation):
\begin{subequations}
\begin{align}
 N_0 &=\bar N_0\bigg[1+\sum_{\vec n,\epsilon}  \big(3e^{3\tilde{\alpha}}F_{|1}^{\vec n,\epsilon}a_{\vec n,\epsilon} - F_{|2}^{\vec n,\epsilon}\big)\bigg]-3e^{3\tilde{\alpha}}\sum_{\vec n,\epsilon}G_{\vec n,\epsilon}a_{\vec n,\epsilon},\\
 g_{\vec n,\epsilon}&= G_{\vec n,\epsilon}-\bar N_0F_{|1}^{\vec n,\epsilon},\\
 k_{\vec n,\epsilon}& =K_{\vec n,\epsilon}- \bar N_0 \bigg(F_{\_1}^{\vec n,\epsilon}-3 \frac{e^{-3{\tilde\alpha}}}{\pi_{\tilde\alpha}} V^{\vec n,\epsilon}_{p_2} + \frac92e^{-3{\tilde\alpha}} V^{\vec n,\epsilon}_{p_3}\bigg),
\end{align}
\end{subequations}
where $a_{\vec n,\epsilon}$ is given in Eq. \eqref{a}.
These formulas define the modes of the lapse function and the shift vector, given in Eqs. \eqref{lapse} and \eqref{shift}, in terms of the variables and Lagrange multipliers of our gauge-invariant formulation, up to terms that are neglected at our truncation order in the perturbations. 

To complete the expressions of the spacetime metric components in our formulation, defined by Eqs. \eqref{eqs:expansions}, as well as of the scalar field \eqref{field}, we just need to substitute $a_{\vec n,\epsilon}$,  $b_{\vec n,\epsilon}$, and $f_{\vec n,\epsilon}$ by their values given in Eq. \eqref{eq:transf}, and to replace $\alpha$ and $\varphi$ in terms of $\tilde\alpha$ and $\tilde\varphi$ [see Eqs. \eqref{alphapp} and \eqref{phiapp}]. Recall that the difference between old and new variables for the homogeneous sector is quadratic in the perturbations, and therefore some contributions arising from the above replacement may be disregarded as negligible within our perturbative truncations (namely, if they provide quadratic or higher order corrections to the inhomogeneous modes, or cubic or higher corrections to the zero-modes).

\section{Negligible second derivatives: self-consistency of the approximation}
\label{appb}

In this appendix we investigate the self-consistency of neglecting the second derivative of the wave function of the MS modes with respect to the homogeneous part of the scalar field in the constraint equation of the Born-Oppenheimer states. 

Suppose that, in Eq. \eqref{constrainBO}, we can ignore  $\langle \hat{\Theta}_{o} \rangle_{\Gamma}$ compared to $\langle \hat{\tilde{\mathcal H}}_0 \rangle_{\Gamma}$. This assumption is justified on the basis of our perturbative scheme. Then, from that equation, we get
\begin{eqnarray}\label{constrainBO2}
\hat{\pi}_{\tilde\varphi}\psi&=&\frac1{2 \langle \hat{\tilde{\mathcal H}}_0\rangle_{\Gamma}}\bigg[\langle \hat{\Theta}_{e}+ \big(\hat{\Theta}_{o} \hat{\tilde{\mathcal H}}_0\big)_S\rangle_{\Gamma}
+i \langle   {\mathrm d}_{\tilde\varphi}\hat{\tilde{\mathcal H}}_0 - \frac{1}{2}{\mathrm d}_{\tilde\varphi}\hat{\Theta}_{o} \rangle_{\Gamma} +\langle \hat{\mathcal H}_0^{(2)} - (\hat{\tilde{\mathcal H}}_0)^2 \rangle_\Gamma\bigg]\psi \nonumber\\
&&-\frac{1}{2 \langle \hat{\tilde{\mathcal H}}_0 \rangle_{\Gamma}}\hat{\pi}_{\tilde\varphi}^2 \psi.
\end{eqnarray}
Now we act on this equation with $\hat{\pi}_{\tilde\varphi}$ and compute the value of the second derivative of $\psi$. Taking into account that $[\hat{\pi}_{\tilde\varphi}, \langle \hat O \rangle_{\Gamma}]=-i  \langle {\mathrm d}_{\tilde\varphi} \hat O \rangle_{\Gamma}$ and eliminating terms which are negligible perturbatively, we arrive at
\begin{align}\label{constrainBO3}
&\left[\frac{3i\langle {\mathrm d}_{\tilde\varphi}\hat{\tilde{\mathcal H}}_0 \rangle_{\Gamma}+\langle \hat{\mathcal H}_0^{(2)} - (\hat{\tilde{\mathcal H}}_0)^2 \rangle_\Gamma}{2\langle \hat{\tilde{\mathcal H}}_0 \rangle_{\Gamma}} +2  \langle \hat{\tilde{\mathcal H}}_0 \rangle_{\Gamma}\right] \hat{\pi}_{\tilde\varphi}^2\psi 
= \frac{2i\langle {\mathrm d}_{\tilde\varphi}\hat{\tilde{\mathcal H}}_0 \rangle_{\Gamma}+\langle \hat{\mathcal H}_0^{(2)} - (\hat{\tilde{\mathcal H}}_0)^2 \rangle_\Gamma}{2\langle \hat{\tilde{\mathcal H}}_0 \rangle_{\Gamma}} \nonumber\\
&\qquad\times\left[2\langle \hat{\Theta}_{e}+ \big(\hat{\Theta}_{o} \hat{\tilde{\mathcal H}}_0\big)_S\rangle_{\Gamma}
+\frac{i}{2} \langle   3{\mathrm d}_{\tilde\varphi}\hat{\tilde{\mathcal H}}_0 - 2{\mathrm d}_{\tilde\varphi}\hat{\Theta}_{o} \rangle_{\Gamma}+\frac5{4}\langle \hat{\mathcal H}_0^{(2)} - (\hat{\tilde{\mathcal H}}_0)^2 \rangle_\Gamma\right] \psi \nonumber\\
&\qquad-i
\left[\langle {\mathrm d}_{\tilde\varphi}\hat{\Theta}_{e}+ {\mathrm d}_{\tilde\varphi}\big(\hat{\Theta}_{o} \hat{\tilde{\mathcal H}}_0\big)_S\rangle_{\Gamma}
+i \langle   {\mathrm d}_{\tilde\varphi}^2\hat{\tilde{\mathcal H}}_0 - \frac{1}{2}{\mathrm d}_{\tilde\varphi}^2\hat{\Theta}_{o} \rangle_{\Gamma} +\langle {\mathrm d}_{\tilde\varphi} \hat{\mathcal H}_0^{(2)} - {\mathrm d}_{\tilde\varphi}(\hat{\tilde{\mathcal H}}_0)^2 \rangle_\Gamma\right] \psi  \nonumber\\
&\qquad -\frac1{8}\frac{(\langle \hat{\mathcal H}_0^{(2)} - (\hat{\tilde{\mathcal H}}_0)^2 \rangle_\Gamma)^2}{\langle \hat{\tilde{\mathcal H}}_0 \rangle_{\Gamma}}\psi- \hat{\pi}_{\tilde\varphi}^3 \psi.
\end{align}

Starting with this equation, it is possible to see by iteration (operating repeatedly with  $\hat{\pi}_{\tilde\varphi}$) whether it is legitimate to assume that the action of $\hat{\pi}_{\tilde\varphi}^{n+1}$ on $\psi$ is negligible compared to the action of $\hat{\pi}_{\tilde\varphi}^{n}$, and in particular if this happens for $n=1$. For this, one can see that it is sufficient that $\langle \hat{\mathcal H}_0^{(2)} - (\hat{\tilde{\mathcal H}}_0)^2 \rangle_\Gamma$ be negligible compared with linear terms in the perturbations, and that $\langle {\mathrm d}_{\tilde\varphi} \hat{\mathcal H}_0^{(2)} - {\mathrm d}_{\tilde\varphi}(\hat{\tilde{\mathcal H}}_0)^2 \rangle_\Gamma$ be negligible as well compared to terms of quadratic order, all this assuming that 
\begin{itemize}
 \item[i)]$\langle {\mathrm d}_{\tilde\varphi}\hat{\Theta}_{e}+ {\mathrm d}_{\tilde\varphi}\big(\hat{\Theta}_{o} \hat{\tilde{\mathcal H}}_0\big)_S\rangle_{\Gamma}$ is negligible with respect to terms of quadratic order in the perturbations,
 \item[ii)]$\langle {\mathrm d}_{\tilde\varphi}\hat{\tilde{\mathcal H}}_0 \rangle_{\Gamma}$ and $\langle {\mathrm d}_{\tilde\varphi}\hat\Theta_o \rangle_{\Gamma}$ are at most of the order of quadratic terms.
\end{itemize}

Let us recall that $-i{\mathrm d}_{\tilde\varphi}\hat O= [\hat{\pi}_{\tilde\varphi},\hat O]-[\hat{\tilde{\mathcal H}}_0,\hat O]$. The first term is a generalized derivative with respect to the explicit dependence on $\tilde\varphi$ of the considered operator $\hat O$. For the relevant operators in our discussion, this dependence comes exclusively from the potential of the matter scalar field, that we have considered to be given by a mass term. Then, if the possible values of the mass are considerably small, the term $\langle[\hat{\pi}_{\tilde\varphi},\hat O]\rangle_\Gamma$ may be treated perturbatively, e.g. by expressing the mass value as a certain power of the amplitude parameter of the inhomogeneous perturbations. On the other hand, when $\hat  O$ is any of the theta-operators, the second term $\langle[\hat{\tilde{\mathcal H}}_0,\hat O]\rangle_\Gamma$ gives a non-vanishing contribution, whose details will depend on the particular representation chosen for the homogeneous geometry, 
 and on the properties of the state $\Gamma$.
In consequence, whether the above conditions are true or not will depend on these particular details and should be carefully checked.\footnote{For the case when the representation of the homogeneous geometry is that of LQC, which we apply in Sec. \ref{sec:lqc}, we refer the reader to the discussion of Ref. \cite{hybridFRWflatMS}.}

\bibliographystyle{plain}

\begin{thebibliography}{99}
			
\bibitem{inflationmodels} A.A. Starobisnky, {\it Spectrum of relict gravitational radiation and the early state of the universe}, JETP Lett. {\bf 30}, 682 (1979); \\
A.H. Guth {\it Inflationary universe: A possible solution to the horizon and flatness problems}, Phys. Rev. D {\bf 23}, 347 (1981); \\
A.D. Linde {\it A new inflationary universe scenario: A possible solution of the horizon, flatness, homogeneity, isotropy and primordial monopole problems}, Phys. Rev. Lett. B {\bf 108}, 389 (1982); {\it Chaotic inflation}, Phys. Rev. Lett. B {\bf 129}, 177 (1983).
			
\bibitem{precision} G. Hinshaw {\it et al.}, {\it Nine-year Wilkinson Microwave Anisotropy Probe (WMAP) observations: Cosmological parameter results}, ApJS {\bf 208}, 19  (2013);\\
P.A.R. Ade {\it et al.} (Planck Collaboration), {\it Planck 2015 results. XIII. Cosmological parameters}, {\tt arXiv:1502.01589}.
			
\bibitem{gravpert} P.A.R. Ade {\it et al.} (BICEP2 Collaboration), {\it BICEP2 I: Detection of B-mode polarization at degree angular scales}, Phys. Rev. Lett. {\bf 112}, 241101 (2014);\\
BICEP2/Keck, Planck Collaborations, {\it A joint analysis of BICEP2/Keck array and Planck data}, Phys. Rev. Lett. {\bf 114}, 101301 (2015). 
			
\bibitem{theorypertu} R. Durrer, {\it Cosmological perturbation theory}, {\tt arXiv:astro-ph/0402129};\\
D. Langlois, {\it Lectures on inflation and cosmological perturbations}, Lect. Notes Phys. {\bf 800}, 1 (2010);\\
V.F. Mukhanov, H.A. Feldman, and R.H. Brandenberger, {\it Theory of cosmological perturbations}, Phys. Rep. {\bf 215}, 203 (1992).
			
\bibitem{inflation} A.R. Liddle and D.H. Lyth, {\it Cosmological inflation and large-scale structure} (Cambridge University Press, Cambridge, U.K., 2000);\\ J. Martin, {\it Inflationary cosmological perturbations of quantum-mechanical origin}, Lect. Notes Phys. {\bf 669}, 199 (2005).
			
\bibitem{EGS} J. Ehlers, P. Geren, and R.K. Sachs, {\it Isotropic solutions of the Einstein-Liouville equations}, J. Math. Phys. {\bf 9}, 1344 (1968);\\
W.R. Stoeger, R. Maartens, and G.F.R. Ellis, {\it Proving almost-homogeneity of the universe -- An almost Ehlers-Geren-Sachs theorem}, Astrophys. J. {\bf 443}, 1 (1995).
			
\bibitem{Lpert} E. Lifshitz, {\it On the gravitational stability of the expanding universe}, Zh. Eksp. Teor. Fiz. {\bf 16},  587 (1946);\\
E. Lifshitz and I.M. Khalatnikov, {\it Investigations in relativistic cosmology}, Adv. Phys. {\bf 12}, 185 (1963).
			
\bibitem{Hpert} S.W. Hawking, {\it Perturbations of an expanding universe}, Astrophys. J. {\bf 145}, 544 (1966). 
			
\bibitem{Olson}	D.W. Olson, {\it Density perturbations in cosmological models}, Phys. Rev. D {\bf 14}, 327 (1976).
					
\bibitem{bardeen} J.M. Bardeen, {\it Gauge-invariant cosmological perturbations}, Phys. Rev. D {\bf 22}, 1882 (1980).
			
\bibitem{gaugeinvariant} G.F.R. Ellis and M. Bruni, {\it Covariant and gauge-invariant approach to cosmological density fluctuations}, Phys. Rev. D {\bf 40}, 1804 (1989).
			
\bibitem{sasa} M. Sasaki, {\it Gauge invariant scalar perturbations in the new inflationary universe}, Prog. Theor. Phys. {\bf 70}, 394 (1983);\\
H. Kodama and M. Sasaki, {\it Cosmological perturbation theory}, Prog. Theor. Phys. Suppl. {\bf 78}, 1 (1984).
				
\bibitem{mukhanov} V. Mukhanov, {\it Quantum theory of gauge-invariant cosmological perturbations} Zh. Eksp. Teor. Fiz. {\bf 94}, 1 (1988); Sov. Phys. JETP {\bf 67}, 1297 (1988); {\it Physical foundations of cosmology}, (Cambrige University Press, Cambridge, U.K., 2005).
			
\bibitem{bunchdavies} T.S. Bunch and P.C.W. Davies, {\it Quantum field theory in de Sitter space: Renormalization by point splitting}, Proc. R. Soc. A {\bf 360}, 117 (1978).
			
\bibitem{HH} J.J. Halliwell and S.W. Hawking, {\it Origin of structure in the universe}, Phys. Rev. D {\bf 31}, 1777 (1985).
			
\bibitem{shiwa} I. Shirai and S. Wada, {\it Cosmological perturbations and quantum fields in curved space}, Nucl. Phys. B {\bf 303}, 728 (1988). 
			
\bibitem{Langlois} D. Langlois, {\it Hamiltonian formalism and gauge invariance for linear perturbations in inflation}, Class. Quantum Grav. {\bf 11}, 389 (1994).
			
\bibitem{PinhoPinto} E.J.C. Pinho and N. Pinto-Neto, {\it Scalar and vector perturbations in quantum cosmological backgrounds}, Phys. Rev. D {\bf 76}, 023506 (2007); \\
F.T. Falciano and N. Pinto-Neto, {\it Scalar perturbations in scalar field quantum cosmology}, Phys. Rev. D {\bf 79}, 023507 (2009).
			
\bibitem{lqc} M. Bojowald, {\it Loop Quantum Cosmology}, Living Rev. Rel. {\bf 11}, 4 (2008);\\
G.A. Mena Marug\'an, {\it A brief introduction to Loop Quantum Cosmology}, AIP Conf. Proc. {\bf 1130}, 89 (2009); {\it Loop Quantum Cosmology: A cosmological theory with a view}, J. Phys. Conf. Ser. {\bf 314}, 012012 (2011);\\
K. Banerjee, G. Calcagni, and M. Mart\'\i n-Benito, {\it Introduction to Loop Quantum Cosmology}, SIGMA {\bf 8}, 016 (2012).
				
\bibitem{lqg} T. Thiemann, {\it Modern canonical quantum general relativity} (Cambridge University Press, Cambridge, U.K., 2007);\\
K. Giesel and H. Sahlmann, {\it Loop Quantum Gravity}, PoS(QGQGS 2011)002.

\bibitem{bounce1} A. Ashtekar, T. Paw\l{}owski, and P. Singh, {\it Quantum nature of the Big Bang: An analytical and numerical investigation}, Phys. Rev. D {\bf 73}, 124038 (2006).

\bibitem{bounce2} A. Ashtekar, T. Paw\l{}owski, and P. Singh, {\it Quantum nature of the Big Bang: Improved dynamics}, Phys. Rev. D {\bf 74}, 084003 (2006).

\bibitem{hybridFRW} M. Fern\'andez-M\'endez, G.A. Mena Marug\'an, and J. Olmedo, {\it Hybrid quantization of an inflationary universe}, Phys. Rev. D {\bf 86}, 024003 (2012);   {\it Effective dynamics of scalar perturbations in a flat Friedmann-Robertson-Walker spacetime in Loop Quantum Cosmology}, Phys. Rev. D {\bf 89},  044041 (2014).

\bibitem{hybridFRWflat} M. Fern\'andez-M\'endez, G.A. Mena Marug\'an, and J. Olmedo, {\it Hybrid quantization of an inflationary model: The flat case}, Phys. Rev. D {\bf 88}, 044013 (2013).

\bibitem{hybridFRWflatMS} L. Castell\'o Gomar, M. Fern\'andez-M\'endez, G.A. Mena Marug\'an, and J. Olmedo, {\it Cosmological perturbations in Hybrid Loop Quantum Cosmology: Mukhanov-Sasaki variables}, Phys. Rev. D {\bf 90}, 064015 (2014). 
			
\bibitem{hybridgowdy} M. Mart\'\i{}n-Benito, L.J. Garay, and G.A. Mena Marug\'an, {\it Hybrid quantum Gowdy cosmology: Combining Loop and Fock quantizations}, Phys. Rev. D {\bf 78}, 083516 (2008);\\
G.A. Mena Marug\'{a}n and M. Mart\'{\i}n-Benito, {\it Hybrid quantum cosmology: Combining Loop and Fock quantizations}, Int. J. Mod. Phys. A {\bf 24}, 2820 (2009);\\
L.J. Garay, M. Mart\'\i{}n-Benito, and G.A. Mena Marug\'an, {\it Inhomogeneous Loop Quantum Cosmology: Hybrid quantization of the Gowdy model}, Phys. Rev. D {\bf 82},  044048 (2010);\\
M. Mart\'\i{}n-Benito, G.A. Mena Marug\'an, and E. Wilson-Ewing, {\it Hybrid quantization: From Bianchi I to the Gowdy model}, Phys. Rev. D {\bf 82}, 084012  (2010);\\
D. Brizuela, G.A. Mena Marug\'an, and T. Paw\l{}owski, {\it Big Bounce and inhomogeneities}, Class. Quant. Grav. {\bf 27}, 052001 (2010).
			
\bibitem{dressed} I. Agullo, A. Ashtekar, and W. Nelson, {\it Extension of the quantum theory of cosmological perturbations to the Planck era}, Phys. Rev. D {\bf 87}, 043507 (2013); {\it The pre-inflationary dynamics of Loop Quantum Cosmology: Confronting quantum gravity with observations}, Class. Quantum Grav. {\bf 30}, 085014 (2013).
			
\bibitem{effective} M. Bojowald, G.M. Hossain, M. Kagan, and S. Shankaranarayanan, {\it Anomaly freedom in perturbative Loop Quantum Gravity}, Phys. Rev. D {\bf 78}, 063547 (2008); {\it Gauge invariant cosmological perturbation equations with corrections from Loop Quantum Gravity}, Phys. Rev. D {\bf 79}, 043505 (2009); Phys. Rev. D {\bf 82}, 109903(E) (2010);\\
M. Bojowald, G. Calcagni, and S. Tsujikawa, {\it Observational constraints on Loop Quantum Cosmology}, Phys. Rev. Lett. {\bf 107}, 211302 (2011);\\
T. Cailleteau, L. Linsefors, and A. Barreau, {\it Anomaly-free perturbations with inverse-volume and holonomy corrections in Loop Quantum Cosmology}, Class. Quantum Grav. {\bf 31}, 125011 (2014).\\

\bibitem{BBC} A. Barrau, M. Bojowald, G. Calcagni, J. Grein, and M. Kagan, {\it Anomaly-free cosmological perturbations in effective canonical quantum gravity}, {\tt arXiv:1404.1018}.
			
\bibitem{bianca} B. Dittrich and J. Tambornino, {\it Gauge invariant perturbations around symmetry reduced sectors of general relativity: Applications to cosmology}, Class. Quantum Grav. {\bf 24}, 4535 (2007).

\bibitem{kk} C. Kiefer and M. Kr\"amer, {\it Quantum gravitational contributions to the cosmic microwave background anisotropy spectrum}, Phys. Rev. Lett. {\bf 108}, 021301 (2012).

\bibitem{gow} A. Corichi, J. Cortez, and G.A. Mena Marug\'an, {\it Quantum Gowdy T3 model: A unitary description}, Phys. Rev. D {\bf 73}, 084020 (2006).
			
\bibitem{uniqueness1} J. Cortez, G.A. Mena Marug\'an, J. Olmedo, and J.M. Velhinho, {\it Criteria for the determination of time dependent scalings in the Fock quantization of scalar fields with a time dependent mass in ultrastatic spacetimes}, Phys. Rev. D {\bf 86}, 104003 (2012).

\bibitem{uniqueness2} J. Cortez, G.A. Mena Marug\'an, J. Olmedo, and J.M. Velhinho, {\it A uniqueness criterion for the Fock quantization of scalar fields with time-dependent mass}, Class. Quantum Grav. {\bf 28}, 172001 (2011).

\bibitem{uniquenessother} L. Castell\'o Gomar, J. Cortez, D. Mart\'\i{}n-de Blas, G.A. Mena Marug\'an, and J.M. Velhinho, {\it Uniqueness of the Fock quantization of scalar fields in spatially flat cosmological spacetimes}, JCAP {\bf 11}, 001 (2012);\\
J. Cortez, L. Fonseca, D. Mart\'{i}n-de Blas, and G.A. Mena Marug\'an, {\it Uniqueness of the Fock quantization of scalar fields under mode preserving canonical transformations varying in time}, Phys. Rev. D {\bf 87}, 044013 (2013);\\
J. Cortez, D. Mart\'{i}n-de Blas, G.A. Mena Marug\'an, and J.M. Velhinho, {\it Massless scalar field in de Sitter spacetime: Unitary quantum time evolution}, Class. Quantum Grav. {\bf 30}, 075015 (2013);\\
L. Castell\'o Gomar and G.A. Mena Marug\'an, {\it Uniqueness of the Fock quantization of scalar fields and processes with signature change in cosmology}, Phys. Rev. D {\bf 89}, 084052 (2014).

\bibitem{uniqueperturb} M. Fern\'andez-M\'endez, G.A. Mena Marug\'an, J. Olmedo, and J.M.  Velhinho, {\it Unique Fock quantization of scalar cosmological perturbations}, Phys. Rev. D {\bf 85}, 103525 (2012).

\bibitem{mmo}  M. Mart\'{i}n-Benito,  G.A. Mena Marug\'an, and J. Olmedo, {\it Further improvements in the understanding of isotropic Loop Quantum Cosmology}, Phys. Rev. D {\bf 80}, 104015 (2009).

\bibitem{pres} G.A. Mena Marug\'an, J. Olmedo, and T. Pawlowski, {\it Prescriptions in Loop Quantum Cosmology: A comparative analysis}, Phys. Rev. D {\bf 84}, 064012 (2011).

\bibitem{Immirzi} G. Immirzi, {\it Quantum gravity and Regge calculus}, Nucl. Phys. Proc. Suppl. {\bf 57}, 65 (1997);
{\it Real and complex connections for canonical gravity}, Class. Quantum Grav. {\bf 14}, L177 (1997).

\bibitem{vidotto} C. Rovelli and F. Vidotto, {\it Stepping out of homogeneity in Loop Quantum Cosmology}, Class. Quantum Grav. {\bf 25}, 225024 (2008). 

\end{thebibliography}

\end{document}